\documentclass[12pt,]{article}
\usepackage[dvips]{graphics,color,epsfig}
\usepackage[square,comma,sort&compress]{natbib}
\newcommand{\diag}{\hbox{diag\,}}
\renewcommand{\Im}{\hbox{Im\,}}
\renewcommand{\Re}{\hbox{Re\,}}

\title{Optics of Nanostructured Fractal Silver Colloids}
\author{$^1$S.V. Karpov\thanks{e-mail: karpov@iph.krasn.ru}, $^1$A.L. Bas'ko,
$^{1,3}$A.K. Popov,
$^{2}$V.V. Slabko\\
and $^3$Thomas F. George\thanks{e-mail: tgeorge@uwsp.edu}\vspace{5mm}\\
$^1$L.V. Kirenskii Institute of Physics, Siberian Division,\\
Russian
Academy of Sciences, Krasnoyarsk, 660036, Russia\\
$^{2}$Engineering-Physical Department,\\ Krasnoyarsk State Technical University, 660028, Russia\\
$^{3}$Office of the Chancellor/ Departments of Chemistry\\ and
Physics \& Astronomy,\\ University of Wisconsin-Stevens Point,\\
Stevens Point, WI 54481-3897, USA\\
\\
\small A chapter in {\it Recent Research Developments in Optics},
vol. 2,\\ \small Research Signpost, Trivandrum, India, 2002\\
\\
Dated: May 14, 2002}
\date{}
\begin{document}
\maketitle
\begin{abstract}
Based on the theory of the optical properties of fractal clusters,
which is an operator-based modification of the coupled-dipole
method, an alternate solution is proposed for the problem of
adequately describing the evolution of optical spectra of any
polydisperse silver colloid with particles falling within the
range of most characteristic sizes ($5 - 30$ nm). This is the
range over which the results of the application of the well-known
methods of classical electrodynamics, including the Mie theory,
disagree with experimental data. The effect of variation of the
parameters of such media on optical spectra is studied by a
numerical simulation, which accounts for particle electrodynamic
dipole-dipole interactions. Indeed, such interactions are shown to
be  a key factor in determining the broadening of the sol
absorption spectra during the course of fractal aggregation. A
quantitative explanation is given for the reasons for the
appearance of individual specific features in the contours of the
spectral absorption of different types of silver colloids.


\end{abstract}
\thispagestyle{empty}
\tableofcontents
\section{ INTRODUCTION}
Since  the famous Faraday's tractates  became available over a century ago,
the question about the origin of the intense color of  sols containing small
noble metal particles has long been one of the most complex and still
unsolved problems in the fields of colloid chemistry and the optics of
dispersion media. The interest is motivated by the fact that optical spectra
provide information about the most practically important physical
characteristics of sols, such as the particle size and  thickness of the
adsorption layer. The spectra enable one to monitor the changes of the
electronic structure of small particles when a transition from the bulk
materials to clusters occurs. On the one hand, the color of such sols is
explained by the true light absorption by the particles and is also affected
by light scattering.   For low-concentrated sols, the role of the latter
factor is insignificant; however, despite this fact, all basic regularities
associated with  the dramatic changes in color are also observed in these
sols \cite{1}.

The most widespread viewpoint about this problem, beginning with the
well-known  work by Mie \cite{2}, is based on the idea of the spectral
selectivity of scattering and absorption, which is governed by the particle
size. According to this work, variations in the optical spectra of noble
metal colloids are associated with the dependence of the position of the
maximum of absorption (scattering) band ($\lambda_r$) on particle size,  and
the appearance of the long-wavelength wing in the absorption spectrum seems
to have resulted from an arbitrary increase in the initial size of the
particles in  solution \cite{2,3,4,5,6}. The restrictions imposed on the
treatment of this effect within the framework of these representations were
mainly reduced to the condition of preserving the spherical shape of the
particles. It is this concept that is described in most current textbooks on
colloid chemistry (see, for example, \cite{7}).

The Mie theory is most often used to interpret the optical spectra of metal
sols, although in this case, the achievement of agreement between calculated
and experimental spectra requires the existence of particles in a sol to
fall within too wide a range of their sizes (\cite{8} and references
therein). Here, the maximal particle size far exceeds the values
characteristic of typical metal colloids. The large discrepancy with the
experiment was explained mainly by the nonsphericity of real particles and
the size dependence of the refractive index of metals \cite{3}. Accounting
for size effects in \cite{5} made it possible to predict a larger (compared
with \cite{4,6}) broadening of the absorption spectra of silver colloids in
terms of the Mie theory. However, the existence of excessively large (100 nm
and above) particles in these systems was still recognized.

The most serious discrepancies between interpretations based on the Mie
theory were mentioned in our observations of the evolution of absorption
spectra of various silver hydrosols containing spherical particles with a
diameter of about $2R_i = 5 - 25$ nm  \cite{1}, where in one of these sols,
particles sizes were varied, whereas in another they were kept constant.
Despite this, the spectral changes in hydrosols were similar in both cases.
The same discrepancies were mentioned in \cite{9}. In this work, the
statistical function of the particle size distribution in gold and silver
hydrosols was carefully monitored with the help of an electron microscope
over the course of sharp broadening of absorption spectra and,
correspondingly, of the variations in sol color. It was demonstrated
\cite{9} that, in this process, the particle size distribution restricted by
the range $2R_i = 5 - 25$ nm remains virtually unchanged. This implies that,
in this work, the role of the variation in particle size can be considered
insignificant. In this connection, it is reasonable to cite data \cite{10}
stating that, for aerosol sediments of silver particles, the wavelength of
the surface plasmon $\lambda_r$ is practically independent of $R_i$ over the
range $2R_i = 3 - 25$ nm. This is also supported by the data reported in
\cite{6}, which shows that, over the range $2R_i = 5 - 30$ nm, the
calculated shift of the resonant wavelength $\Delta\lambda_r (2R_i )$ is
only 15 nm, whereas the experimentally observed shift in aggregating silver
hydrosols, having the aforementioned particle sizes, is above 400 nm.

The dependence of $\lambda_r$ on  $R_i $ over the range of smaller particle
sizes $2R_i <5$ nm was demonstrated in \cite{5,6,11}; moreover, for some
particle sizes, this dependence can be reversed. However, the absolute
values of the $\Delta\lambda_r (2R_i )$ shifts are also small. The problem
of the size dependence of the frequency of resonant absorption is reviewed
in detail in \cite{9,10}. In general, the authors of these reviews note that
many theoretical and experimental results often contradict each other, thus
confirming that adequate theories are still absent in this field of
research.

In the background of the unresolved controversies regarding the optics of
dispersion media, a new approach to describe optical properties of colloids
has been initiated by  Shalaev, Stockman, Markel, {\it et al.} \cite{12,13}.
In these works, the theory of optical properties of fractal clusters (OPFC)
is provided, which accounts exactly for particle electrodynamic
interactions. In accordance with this theory, the main reason for the
broadening of spectra of colloids is the particle aggregation. Precisely the
same conclusion on the role of aggregation was drawn as far back as a half
of century ago by a number of researchers (for example, see \cite{14}). The
same viewpoint is shared by the authors of the aforementioned work \cite{9},
and only the absence of a theoretical basis did not allow them to provide a
substantial explanation of the results obtained. This idea is favored in a
number of recent publications, in particular in \cite{15,16}.

The appearance of optical coupling of the particles in the aggregated sols
is most pronounced at the Fr\"olich frequency, associated with the
lowest-order surface mode. Over  recent decades, considerable progress has
been made in the development of theoretical models of optical absorption by
the aggregated sols  based on the exact solution of the electrodynamical
problem associated with coupled spherical particles
\cite{17,18,19,20,21,22}. However, because the exact solution of the problem
is difficult to find, a more simple method of coupled dipoles
\cite{23,24,25,26} is often used based on the approximation of the particles
by discrete interacting dipoles \cite{27,28,29}. Note that basic model
concepts about the outlined methods were formulated by De Voe (\cite{30,31})
and employed for calculations of the polarizabilities of the molecular
aggregates. A very effective operator approach to the method of coupled
dipoles was developed by Shalaev, Stockman and Markel \cite{13,32} (see also
the review in \cite{33} and references therein). These authors were the
first to apply the method of coupled dipoles to the colloid metal sol
aggregates. The principal importance of accounting for the fractal geometry
of these aggregates was revealed, and various models and approximations were
developed including an exact theory of the optical properties of fractal
clusters. The application of this method to silver colloids enabled an
increased of the accuracy of the description of their spectra to a level not
achievable with  alternative methods.

It is known that particle aggregation in colloids is accompanied
by the formation of fractal structures, which are assembled from
the main portion of the initially-isolated particles in the
process of sol evolution (see \cite{1}). Only the fractal approach
to describing sol properties made it possible to gain new insight
into well-known facts not restricted just to the optical
properties of fractal nanostructures (for example, see
\cite{32,33}). In particular, it was confirmed  \cite{12,13,32,33}
that some features of the strong influence of neighboring
particles are observed in the spectra of any particles comprising
fractal aggregates (or fractal clusters). This leads to a
noticeable shift in the frequency of the intrinsic optical
resonance of particles ($\omega_r$). The reason for the shift of
resonant frequencies is associated with the dipole-dipole
interaction between the light-induced (oscillating) dipole moments
of each particle and the particles of the surrounding medium (with
a dominating effect of the nearest particle). Since fractal
objects do not possess translational invariance, they cannot
transmit travelling waves. For this reason, optical dipole
excitations in fractals tend to be localized in single particles,
which is why different parts of a fractal (different particles)
absorb light independently. This feature is responsible for the
light-induced modification of fractal aggregates discovered in
\cite{34}. In approximating the pair interaction of particles,
which allows us to explain clearly the essence of the theory of
OPFC, the value of the frequency shift of intrinsic resonance of
the $i$th particle under the effect of the $j$th particle (without
exact account for the total contribution of the far-spaced
particles) is inversely proportional to the third power of the
distance between their geometrical centers ($\Delta\omega_r\propto
R_{ij}^{-3}$). We emphasize that it is precisely the interparticle
distance that is the crucial parameter in the theory of OPFC.

In the present work, it will be shown that all the observed specific
features of the adsorption spectra of silver colloids, as well as the
reasons for the appearance of the long-wavelength wing, can be adequately
explained in terms of the theory of OPFC, where the unique correlation
between the structural and optical properties of fractal aggregates is
expressed. The applicability of this approach, in addition to the
aforementioned condition of the preservation of the spherical shape by the
particles, is also restricted by the smallest admissible diameter $2R_i$,
because this value ($2R_i <2-4$ nm) begins to determine the homogeneous
width of the spectrum of the surface plasmon $\Gamma_i$ due to relaxation
effects at the particle surface \cite{10,11,35}:
\begin{equation}\label{1.1}
\Gamma_i =\Gamma_{bulk} +v_f /R_i ,
\end{equation}
where $\Gamma_{bulk}$, is the relaxation constant of free electrons for a
bulk silver specimen, and $v_f$ is the Fermi velocity. In this case, the
$v_f /R_i$ term can become larger than $\Gamma_{bulk}$. Evidently,
$\Gamma_i$ should be, at least by several times, smaller than the width of
the visible part of the spectrum $\Gamma_v$, because at $\Gamma_i
/\Gamma_v>0.5-1$ no visually registered spectral changes related to the
aggregation in sols with such small particles can be observed, and the sols
will preserve their gray color.

The purpose of this  of this work is to carry out a detailed quantitative
analysis of the effect of dipole-dipole interactions of the particles on the
absorption spectrum of the colloids during the course of their aggregation
into fractal structures. The analysis is performed through a numerical
solution of the equations of the OPFC theory for a polydisperse ensemble of
the coagulating particles.

\section {MATHEMATICAL MODELS OF THE GROWTH OF FRACTAL AGGREGATES} \setcounter{equation}{0}

In this work, to generate fractal aggregates, we used a 3-D model of
cluster-cluster aggregation with the probability of irreversible coagulation
of particles upon their collisions equal to 1. This condition decreases the
counting time, with no effect on the value of the fractal dimension. As a
rule, the total number of particles did not exceed 50 (this restriction was
determined by the calculation time of the optical spectra). At the initial
time, particles whose sizes  fall within the range of $5 - 25$ nm were
uniformly distributed over the space with volume $L^3$ ($L = 200$ nm) with
an arbitrary selection of the direction of motion. The values of the initial
velocities corresponded to the Maxwell distribution. A mean free path
corresponding to the time of motion with no collision was introduced for
each particle. Upon the collision and coagulation of particles, their
intrinsic kinetic energy is transformed into the kinetic energy of
translational and rotational motion of an aggregate. Under the regime of
Brownian aggregation, the mass centers of the assembling aggregates move
along broken linear trajectories. Two models of aggregation (spontaneous
Brownian aggregation of electrically neutral particles and Coulombic
aggregation of initially-charged particles with bipolar charge values of
$\pm 25$ e, where e is the electron charge) were studied in this work. The
existence of a particle charge is associated with the mechanism of mutual
charging in a system \cite{36}. In the case of mutual charging, the
aggregation occurs due to both the short-range van der Waals interaction
($E_w\propto r^{-6} $), whose radius is determined by the condition $E_w >
kT$, and to the long-range Coulombic interaction ($E_c\propto r^{-1}$ ).

The developed algorithm also allows us to calculate various characteristics
of the forming aggregates, i.e., fractal dimension and  degree of
aggregation of a medium determined by the broadening of an absorption
spectrum \cite{37}; to study the kinetics of aggregation as a function of
the viscosity of a dispersion medium, the value of the particle charge and
its sign, as well as of the characteristics of the incident light; to
perform calculations with various functions of the particle size
distribution (FPSD); and to change the position and width of the intrinsic
particle resonance, particle bulk concentration, etc.

The value of the bipolar charges ($\pm25$ e) of the particles was
determined by the condition of the excess of Coulombic interaction
energy between particles separated by a mean distance (typical of
a real hydrosol) over the energy of thermal motion (k$T$).  The
calculated values of the fractal dimension
 of an aggregate for two regimes of aggregation are
equal to $D = 1.78$ for Brownian aggregation and $D = 1.65$ for
Coulombic aggregation, which is attributed to the differences in
the kinetics of the aggregate growth. The half-time of aggregation
\cite{7,37} in the first case was $20 - 25$ times higher than in
the second case.

\section{ SIMULATION OF OPTICAL SPECTRA OF FRACTAL AGGREGATES}
\setcounter{equation}{0}
\subsection{Basic equations}
Unlike the works employed the binary approximation, the algorithm developed
in this work is based on the complete set of equations of the theory of OPFC
\cite{13}, where the authors considered the fractal composed of $N$
particles (with dipole-dipole interactions at optical frequencies) polarized
by the external field ${\bf E}_i^{(0)}$ and located at points ${\bf r}_i$.
Then the dipole moments $d$ induced on different particles obey the system
of equations
\begin{equation}\label{3.1}
d_{i\alpha}=\chi_0E_{i\alpha}^{(0)}-\chi_0 \sum_\beta\sum_{j=1,j\ne i}^N
\frac{\delta_{\alpha\beta}-3n_\alpha^{(ij)}n_\beta^{(ij)}}
{r_{ij}^3}d_{j\beta},
\end{equation}
where $i,j=1,2,3...N$, $\alpha,\beta=\{x,y,z\}$, $\chi_0$  is the dipole
polarizability of a single particle, ${\bf r}_{ij} ={\bf r}_{i}-{\bf
r}_{j}$, and ${\bf n}^{(ij)} ={\bf r}_{ij}/r_{ij}$. If the sizes of the
fractal aggregates are much smaller than the wavelength of an incident beam,
the external field ${\bf E}_i^{(0)}$ at the location of the $i$th particle
can be considered as uniform and independent of $i$. In this case, the
dipole moment induced on the $i$th particle is expressed via the
corresponding linear polarizability $\chi_{\alpha\beta}^{(i)}$ in the
following form:
\begin{equation}\label{3.2}
d_{i\alpha}=\sum_\beta\chi_{\alpha\beta}^{(i)}E_\beta^{(0)}
\end{equation}

The problem consists in determining $\chi_{\alpha\beta}^{(i)}$, because its
imaginary part uniquely determines the light absorption by the $i$th
particle. As seen from (\ref{3.2}), for this purpose it is necessary to find
$d_{i\alpha}$ that solves the system of (\ref{3.1}) with respect to this
parameter.  The solution to this system is performed in a matrix form. To
this end, it is necessary to introduce the matrix $W$ with elements
$$
\langle i\alpha|W|j\beta\rangle=
\frac{\delta_{\alpha\beta}-3n_\alpha^{(ij)}n_\beta^{(ij)}}
{r_{ij}^3},\, i\ne j; \qquad \langle i\alpha|W|i\beta\rangle=0.$$
This matrix acts in the $3N$-dimensional space of the vectors
${\bf d }$ and, ${\bf E}^{(0)}$ with components $d_{i\alpha}$ and
$E_{i\alpha}^{(0)}$. Upon introducing the new complex variable $z$
with real and imaginary parts $-X$ and $-\delta$,
\begin{equation}\label{3.3}
z\equiv -(X+i\delta)=\chi^{-1}_0,
\end{equation}
the main system of (\ref{3.1}) acquires the following form:
\begin{equation}\label{3.4}
(z+W){\bf d}={\bf E}^{(0)}.
 \end{equation}
Because the matrix $W$ is symmetrical, it is reduced to the diagonal form
via an orthogonal transformation,
\begin{equation}\label{3.5}
UWU^{T}=\diag(w_n),  \qquad UU^{T}=1,
\end{equation}
where $\diag(w_n)$ is the diagonal matrix of eigenvalues of $w_n$, and  $U$
is the matrix whose columns are the components of the eigenvectors of $W$
(the superscript $T$ denotes a transposition).

The dipole moment $d$ is determined with the help of (\ref{3.5}) as
\begin{eqnarray}\label{3.6}
W&=&U^{T}\diag(w_n)U,\nonumber\\
z+W&=&U^{T}\diag(z+w_n)U,       \\
(z+W)^{-1}&=&U^{T}\diag(z+w_n)^{-1}U,\nonumber
\end{eqnarray}
and with allowance for (\ref{3.4}) we obtain
    \begin{equation}\label{3.7}
{\bf d}=U^{T}\diag(z+w_n)^{-1}U{\bf E}^{(0)}.
\end{equation}
Thus, expressing the desired values of $d_{i\alpha}$ via the eigenvalues of
$w_n$, and components of the corresponding intrinsic vectors
$U_{i\alpha}^n$, we find
\begin{equation}\label{3.8}
d_{i\alpha}=\sum_\beta\sum_{j=1}^{N}\sum_{n=1}^{3N}
U_{i\alpha}^nU_{j\beta}^n(z+w_n)^{-1}E_{j\beta}^{(0)}.
 \end{equation}

Comparing (\ref{3.8}) and (\ref{3.2}), we obtain the expression
$\chi_{\alpha\beta}^{(i)}$
\begin{equation}\label{3.9}
\chi_{\alpha\beta}^{(i)}=\sum_{j=1}^{N}\sum_{n=1}^{3N}
U_{i\alpha}^nU_{j\beta}^n(z+w_n)^{-1},
\end{equation}
where $U_{i\alpha}^n$ and  $w_n$ are the eigenvalues of the vector and
matrix $W$, respectively. The linear polarizability of the particle in an
aggregate averaged over the particle number $N$ is equal to
$\chi_{\alpha\beta}=N^{-1}\sum\limits_i \chi_{\alpha\beta}^{(i)}$.
 Using symmetry with
respect to rotations and the averaging over the entire aggregate
orientations, the polarizability tensor is reduced to the following scalar:
\begin{equation}\label{3.10}
\chi=\frac13\sum_\alpha\chi_{\alpha\alpha}=
\frac1{3N}\sum_{\alpha}\sum_{i=1}^{N} \sum_{j=1}^{N}\sum_{n=1}^{3N}
U_{i\alpha}^nU_{j\alpha}^n(z+w_n)^{-1}.
\end{equation}
Because the eigenvalues  of $w_{n}$,  and components  of the eigenvectors
$U_{i\alpha}^{n}$ have real values, $z$ becomes the only complex number in
this expression. Substituting  (\ref{3.3}) into (\ref{3.10}), we obtain the
following expression for the function $\Im\chi$ describing the absorption
spectrum of the fractal aggregate:
\begin{equation}\label{3.11}
{\Im}\chi(X)=\frac1{3N}\sum_{\alpha}\sum_{i=1}^{N}
\sum_{j=1}^{N}\sum_{n=1}^{3N}
U_{i\alpha}^nU_{j\alpha}^n\frac{\delta}{(-X+w_n)^2+\delta^2}.
 \end{equation}
\subsection{Dipole polarizability of a two-level system}
If a particle has its resonance at a frequency $\omega_r$, then for a dipole
moment of the transition $d_{12}$ and relaxation rate (homogeneous
halfwidth) $\Gamma$, in the simplest case of a single resonance, the dipole
polarizability of a particle can be described by the expression \cite{12,32}
\begin{equation}\label{3.12}
\chi_0=\frac{|d_{12}|^2}{\hbar(\Omega+i\Gamma)},
\end{equation}
where $\Omega$ is the detuning of resonance, and $\hbar$  is Plank's
constant.  Comparing (\ref{3.12}) and (\ref{3.3}) yields expressions for $X$
and $\delta$ as
\begin{equation}\label{3.13}
X=\frac{\hbar\Omega}{|d_{12}|^2},\quad\delta=\frac{\hbar\Gamma}{|d_{12}|^2}.
 \end{equation}
Here, $X$ has the meaning of relative frequency detuning, and $\delta$
determines the resonance width and amplitude. The expressions of
(\ref{3.13}) allow one, at a fixed value of $d_{12}$, to obtain the
functional dependence ${\Im}\chi(X)$, and hence the absorption spectrum of
an arbitrary aggregate with particle pair interactions  at various values of
its fractal dimension under conditions of linear response  to the external
field.
\subsection{Polarizability with account for the frequency dependence of
the permittivity of a material} The dipole polarizability of a sphere with
radius $R_n$ corrected for the irradiation reaction can be described in
greater detail by the expression reported in \cite{32,33} with allowance for
the spectral dependence of the optical constants of a material comprised by
the particles,
\begin{equation}\label{3.14}
\chi_0=R_n^3\frac {(\epsilon-\epsilon_h)}
{\epsilon+2\epsilon_h-i(2/3)(kR_n)^3(\epsilon-\epsilon_h)},
\end{equation}
where  $\epsilon=\epsilon'+i\epsilon''$ is the permittivity of the material
(in this case, silver), $\epsilon$ is the permittivity of a disperse medium
(water in the case of hydrosols, where $\epsilon_h$ =1.78 in the spectral
region from 200 to 1000 nm), and $k$ is the wavenumber. Expressions for
parameters $\delta={\Im}\chi_0^{-1}$ and $X={\Re}\chi_0^{-1}$  introduced
into formulas (\ref{3.3}) are represented in the following forms:
\begin{equation}\label{3.15}
\delta=\frac {3R_n^{-3}\epsilon''\epsilon_h}
{|\epsilon-\epsilon_h|^2}+2k^3/3,
\end{equation}
\begin{equation}\label{3.16}
X=-R_n^{-3}{[}1+\frac {3\epsilon_h(\epsilon'-\epsilon_h)}
{|\epsilon-\epsilon_h|^2}{]}.
\end{equation}
The permittivity of metal introduced into these expressions can be described
 by the Drude formula,
 \begin{equation}\label{3.17}
\epsilon=\epsilon_0-\frac{\omega_p^2}{\omega(\omega+i\Gamma)},
\end{equation}
where $\epsilon_0$ accounts for the integral contribution to the
permittivity of interband transitions, and $\omega_p$ is the plasma
frequency.

\section{RESULTS AND DISCUSSION} \setcounter{equation}{0}
\subsection{Pair interactions}
In order to demonstrate the effect of the dipole-dipole
interactions of the particles, Fig. 1 shows a set of curves
calculated with the aid of the developed algorithm, which
describes the main regularities of the evolution of the absorption
spectrum of a pair of approaching particles as a function of the
interparticle distance over the range $R_{ij} = 10 - 18$ nm (the
initial spectrum corresponds to the surface plasmon in silver
hydrosols). As has been shown in particular in \cite{13}, the
effect of a neighbor particle is manifested in the splitting of a
characteristic single resonance $\omega_r$ (curve 11) of an
initially isolated single particle and in the appearance of two
peaks: low-frequency $\omega_l$ and high-frequency $\omega_h$,
with the ratio of their frequency shifts with respect to
$\omega_r$ equal approximately to 1/2.
\begin{figure*}[!h]
\begin{center}
\includegraphics[height=.5\textwidth]{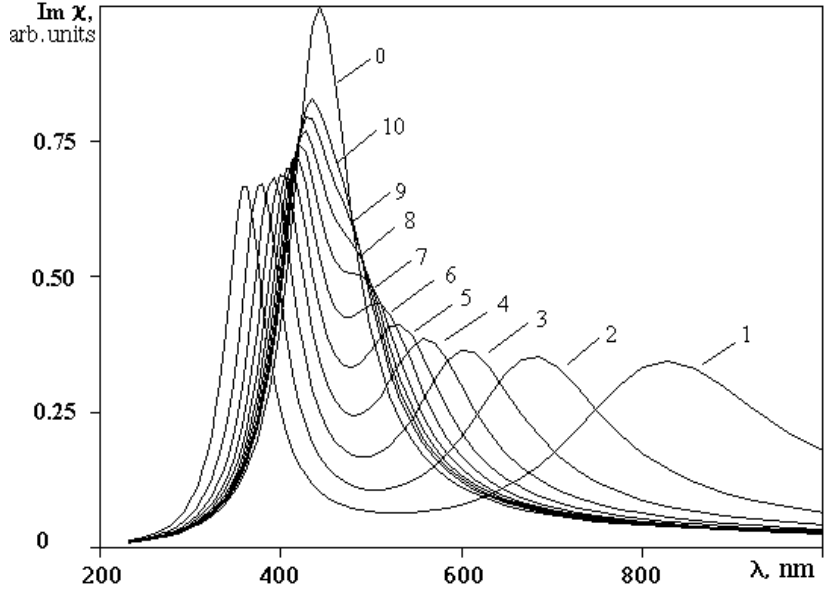}
\caption[recon]{\label{f1} Absorption spectra of  pairs of
particles (with a single identical resonance) determined by their
dipole-dipole interactions as a function of the distance $R_{ij}$
between the particles: (1) 10, (2) 11, (3) 12, (4) 13, (5) 14, (6)
15, (7)16, (8) 17, (9) 18, and (10) 19 nm. Curve 11 corresponds to
the initial spectrum of single non-interacting particles. }
\end{center}
\end{figure*}
As seen from Fig. 1, at an interparticle distance exceeding 20 nm,
spectral changes associated with pair interactions virtually
vanish. At distances of less than 10 nm, the low-frequency peak
approaches the long-wavelength boundary of the optical range and,
upon further decrease in $R_{ij}$, falls outside the limits of
this boundary. As $R_{ij}$ decreases, a broadening of the
low-frequency peak is also observed.

In the presented set of curves, attention should be paid to a rather large
spectral shift of a low-frequency peak observed over the relatively-narrow
range of variation in interparticle distances. We emphasize that the least
possible values of these distances, $R_{ij}^{min} =R_i +R_j$, are limited by
the sizes of the contacting particles, $2R_i$ and $2R_j$.

The fact observed in this approach, that the value of the frequency shift of
the resonant absorption of interacting particles is much more sensitive to
the variation in their sizes, is worth special mention. This is explained by
the higher power of such a dependence. For example, in the theory of OPFC
(pair approximation), the value of the resonant shift of contacting
particles is equal to \hbox{$\Delta\omega_r\propto(R_i+R_j)^{-3}$}, whereas
the Mie theory results, according to data reported in \cite{5,6} for silver
sols, show an almost linear dependence of the value of the frequency shift
of the spectral maximum of absorption on particle size (at least over the
size range $2R_i =20-100$ nm). In addition, an increase in the particle size
leads, according to the theories of OPFC and Mie, to opposite spectral
effects.

During the course of aggregate growth, the diversity of variants of the
spatial environment of each specific particle increases, and the relative
fraction of the particles brought into most intimate contact rises. This
event is accompanied by a gradual increase in the extension of the
long-wavelength wing of the spectrum.

The appearance of the short-wavelength wing of the spectrum of the fractal
aggregate with a twice narrower bandwidth (at a frequency scale) is
attributed to the appearance of a high-frequency peak for the interacting
pairs (Fig. 1). However, the precise experimental registration of the
shortwavelength broadening attributed to the surface plasmon of silver
hydrosols is somewhat complicated due to the superposition of the band wing
of interband absorption within the range $\lambda < 350$ nm and partial
light absorption by a dispersion medium in some sols. Hence, the fractal as
a system composed of $N$ particles is an ensemble of $N$ high-$Q$ resonators
(coupled dipoles) corresponding to optical resonances of approaching
particles with randomly-distributed eigenfrequencies $(\omega_r)_i=\omega_r
-(\Delta\omega_r)_i$. Within the framework of this concept, the fractal
aggregate can also be considered as a statistical set of
arbitrarily-oriented pairs of interacting particles differing in their
positions with respect to each other. In this case, the probability of
existence of arbitrary pairs of particles with the relative interparticle
distance $R_{ij}$ in a fractal obeys the correlation function $g(R_{ij}
)=(D/4\pi)R_0^{-3}(R_{ij} /R_0 )^{D-3}$. The scaling form of the dependence
of the number of particles $N$ in a fractal falling within the sphere of
radius $R_c$ is $N =(R_c /R_0 )^D$, where $R_0$ is a constant corresponding
to the characteristic distance between the nearest particles (scaling length
unit), and $D$ is the fractal (Hausdorff) dimension. Indeed, these
regularities distinguishing fractal objects from disordered systems are
responsible for the appearance of qualitatively-new fractal physical
properties (for an explanation and references see, for example,
\cite{1,33}).

The refined variant of the theory of OPFC, where the effect of the exact
resonance positions of both the nearest particles (pair approximation) and
all other fractal particles are taken into account, was discussed in
\cite{13} (see also the review in \cite{18}. The effect of other particles
was only accounted for in the pair approximation with the aid of the Lorentz
field \cite{12}.
\begin{figure*}[!h]
\begin{center}
\includegraphics[height=.5\textwidth]{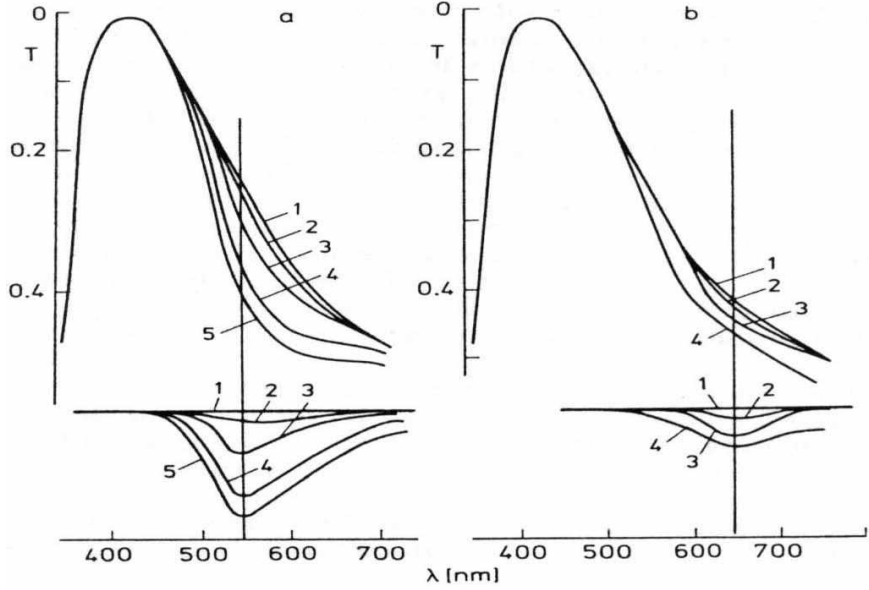}
\caption[recon] {\label{f2}  Transmission spectra (upper) and
spectral dependence of the difference  in absorption (lower) of
the nonirradiated (curve 1) and irradiated ($\tau_{las}\simeq 30$
ps, $W=2\times 10^{-3}$ J/cm$^2$) samples (the silver aggregates
are fixed in gelatin). (a) $\lambda = 540$ nm; curves 2, 3, 4, 5
correspond to 1, 20, 80, 230 pulses; (b) $\lambda = 641$ nm;
curves 2 and 3 correspond to 20 and 120 pulses, respectfully, and
curve 4  to 120 pulses at $W = 8\times 10^{-3}$ J/cm$^2$. }
\end{center}
\end{figure*}
The inhomogeneous  character of spectrum broadening, as well as
other postulates of the theory of OPFC, have been experimentally
confirmed in \cite{1,33,34,38,39}. The  photomodification of
fractal aggregates under the influence of strong laser radiation
proves the inhomogeneous broadening of their absorption spectra.
Let us consider this effect  on the linear optical properties of
fractal aggregates. Figures 2(a-b) show the transmission and
absorption spectra of aggregates before and after irradiation by a
series of strong laser pulses ($\lambda = 540$ nm, $\tau = 30$
sec). One can see that in the latter case, dips appear near the
laser wavelength. The dip width is close to the absorption
linewidth of separate particles in a non-aggregated hydrosol and
is only a small fraction of the aggregate absorption band, i.e.,
the photomodification is selective over the wavelength. With the
increasing number of pulses or growing pulse power, the dip
becomes deeper and broader. Note that a dip was also burnt around
$\lambda = 641$ nm with the radiation obtained by stimulated Raman
scattering  in acetone. Selective modification of the absorption
spectrum of the light in aggregates is observed only at identical
polarizations of the light and the laser radiation. For orthogonal
polarization of the probe beam, no dip is observed in the
absorption spectrum. An increase in the energy density of the
laser beam leads to lower spectral and polarization selectivity.

It has been found that dip burning is a threshold effect in the laser pulse
energy. For ultrashort light pulses, a dip in the absorption spectrum of the
aggregates fixed in gelatin was observed at $W\geq1.5$ MJ/cm$^2$ ($\lambda =
540$ nm). At $\tau = 10$ ns, in order to detect the dip, somewhat higher
energies were requited compared to $\tau = 30$ ps. The spectral width of the
dip increased with the pulse duration growing from 30 ps to 10 ns.

The results obtained support the basic outcomes of the OPFA theory. The
optical response of fractals, despite the long-range dipole-dipole
interaction, is of local character, which allows local modification of the
aggregate.

At a fixed detuning from the resonance of an isolated particle, the
radiation selects only those particles for which the detuning is compensated
by the shift due to the dipole-dipole interaction. Moreover, the
linearly-polarized light "chooses" quite a definite geometrical
configuration of mutual positions of the pairs of particles. This allows, at
a fixed laser frequency, to detect two independent absorption dips,
corresponding to orthogonal polarizations.

At the energy density $W = 1.5\times10^{-3}$ J/cm$^2$  and the particle
concentration $N_0\simeq10^{12}$ cm$^{-3}$, according to estimations, each
"resonant" particle of the aggregate (with a protein adsorption layer)
absorbs about $3\times10^5$ photons per pulse, which corresponds to an
energy sufficient to evaporate a silver particle of 10 nm radius. The area
of energy localization depends on the pulse duration and,  at $\tau\simeq30$
ps, corresponds to the size of one particle, whereas for $\tau\simeq10$ ns
it includes already several particles. This accounts for the high energy
requirements at $\tau\simeq10$ ns and for the partial loss of selectivity.

It should be noted that the width and  shape of the dip depend on the manner
in which aggregates are prepared, the solvent  medium and, in some cases, on
the time interval between irradiation and spectral measurements \cite{39}.
The last dependence indicates relaxation of the structure of photomodified
aggregates. A record of five spectral dips in the visible and near-IR ranges
within the long-wavelength wing of an aggregated Ag sol irradiated with
tunable nanosecond laser radiation was reported in \cite{40}. The properties
of the burnt spectral holes (the color of the irradiated spot) in the
samples of the polymer films contained silver fractal aggregates, which we
investigated, remain very stable already over  $12 - 14$ years (that
promising for dense information recording).
\begin{figure*}[!h]
\begin{center}
\includegraphics[height=.5\textwidth]{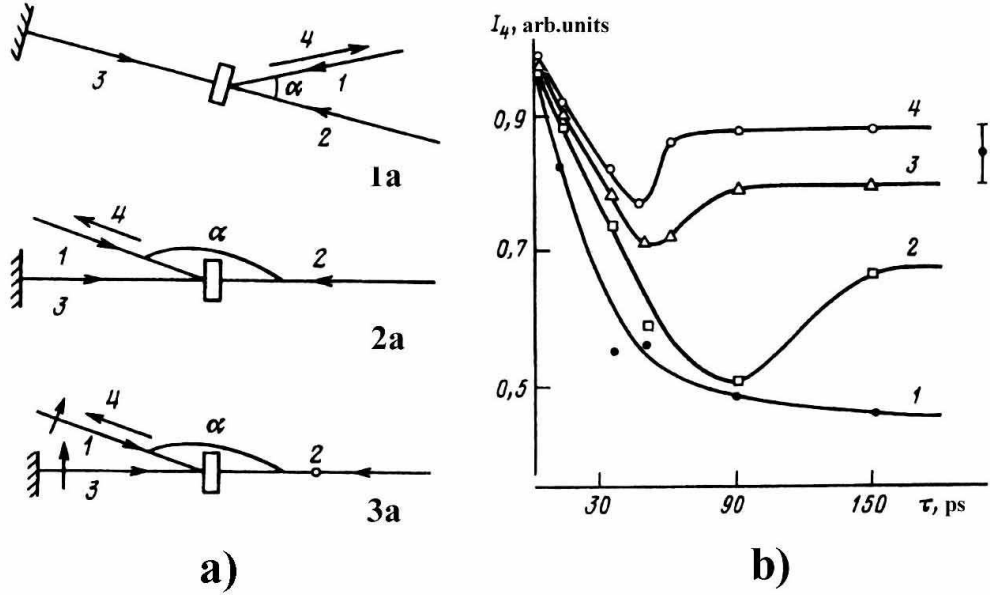}
\caption[recon] {\label{f3}  Various schemes of optical
phase-conjugation experiments with the aggregated silver hydrosol
(a) and the temporal behavior of the photomodification process in
the field of laser pulses (b). Plots 1 - 4 display the energy of
the generated signal 4 vs time delay of the probe pulse 3 with
respect to the overlapping nearly equally intense  30-ps pulses 1
and 2 which produce optical grating in the sol. The delay is
controlled by the shift of the mirror.  The plots 1, 2, 3 and 4
are obtained at intensities of the pulse 2 of $2.75\times
10^{8},\quad 6.92\times 10^{8},\quad 8.25\times 10^{8}$ and
$10^{9}$ W/cm$^2$, correspondingly. }
\end{center}
\end{figure*}
Figure 3 (a,b) displays various experimental schemes (a) and  temporal
behavior of the photomodification process induced by laser pulses $\lambda =
540$ nm, $\tau = 30$ ps at different values of the pulse intensities (b)
\cite{41}. These dependencies have been obtained by the four-wave mixing
(optical phase conjugation) technique through a delay of the probe pulse.
Simultaneous pulses 1 and 2 [Fig. 3(a)] cause photomodification of the Ag
fractal aggregates in a hydrosol cell. A delay of the probe pulse 3, which
length is about 1 cm, is controlled by the shift of the mirror. It  scatters
on the optical grating produced in such a way inside the Ag sol, and
generated signal 4 is recorded vs time delay of the probe pulse 3 with
respect to the photomodification-stimulating pulses 1 and 2. Plot 1
corresponds to a relatively low intensity of the stimulating waves. It
displays instant electronic response, which then decays. The plots 2, 3 and
4 are obtained at higher pulse intensities. They  reveal strong contribution
of the grating produced through the photomodification and display the
dynamics and the delay of this process. Here we see that the higher the
intensity, the faster is the process of photomodification.

A negligible role of the thermal processes was proved through change of the
grating period, which is proportional to $1/sin(\alpha/2)$ [Fig. 3 (a)]. The
angle $\alpha$ was 5$^o$ in the scheme 1a and 160$^o$ in the scheme 2(a).
The estimates and direct measurements revealed that the time interval,
required for the formation of the thermal grating, is longer than 200 ps.
Investigation of polarization effects confirmed these conclusions (see
scheme 3(a), where polarization of waves 1 and 2 are orthogonal).

\subsection{Specific  features of the absorption
spectra of silver colloids}
\subsubsection{Preparation of Ag  colloids}
Different methods were used to prepare colloids (hydrosols):

1. The boronhydride method described in \cite{9,39} proceeds as follow. $1 -
3$ mg of sodium boronhydride is dissolved in 20 ml of cooled bidistilled
water, and  5 ml of such water is used to dissolve 2 mg of silver nitrate.
Then the silver nitrate solution is quickly added to the test tube with
sodium boronhydride solution, and the mixture is intensively shaken. The
resultant colloid is yellow-colored. The extinction spectrum of a fresh
hydrosol has a peak at $\lambda  = 410 - 420$ nm with FWHM in the range $60
- 80$ nm in various measurement. The resonant width is  larger than that of
isolated particles. In particular, this is believed to be due to dispersion
in the size and shape of the particles leading to small inhomogeneous
broadening of the extinction spectra of non-aggregated hydrosols.

2. The second utilized method is based on the use of collargol \cite{1,39},
which a mixture of silver with proteins that stabilize hydrosols
(C-hydrosol). 1 ml of collargol dissolved in 10 ml of bidistilled water
yielded an orange-brown solution, in its extinction spectrum  a peak was
observed at $\lambda $ = 420 nm, broadened towards to the long-wavelength
wing. The broadening is believed to be caused by the silver particles
combining with protein molecules to form a complex wherein particles are
spaced at distances comparable or somewhat larger than their diameters ($10
- 15$ nm). Stabilized isolated particles, showing no changes in the
absorption spectra within a month period, were prepared by heating of the
collargol solution with a small addition of sodium nitrate. Aggregation of
the hydrosol obtained was initiated by adding 0.1 NaOH solution in the
proportion 1:10. In 1 - 2 weeks the hydrosol became dark-red, and its
spectrum displayed a high long-wavelength wing.

3. The third method is based on the reduction of silver by ethyl alcohol.
The colloid prepared according to this method is called A-hydrosol. The
preparation of 100 ml of an A-hydrosol requires 0.4 g of AgNO$_3$ and 0.3 g
of PVP. The reduction of silver was conducted by heating this A-hydrosol in
a solution consisting of 20 ml of H$_2$O and 80 ml of C$_2$H$_5$OH for 20
min in the range $348 - 353$ K. Partial aggregation of the solution during
its heating depended on the heating time, which varied over the range $3 -
50$ min.

\subsubsection{Evolution of the spectra of silver hydrosols during
spontaneous aggregation of the disperse phase}

{\begin{figure*}[!h]
\begin{center}
\includegraphics[height=.5\textwidth]{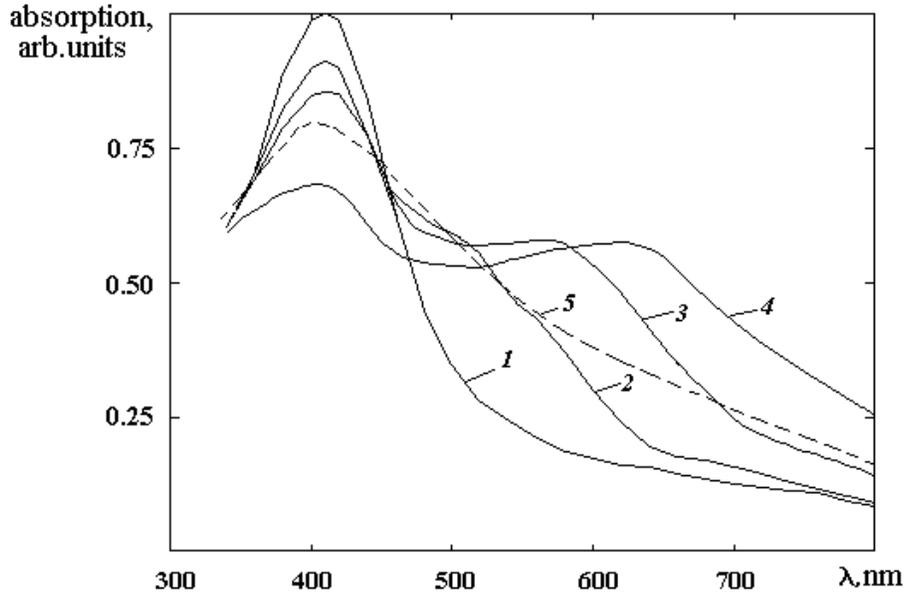}
\caption[recon] {\label{f4}Typical absorption spectra of silver
hydrosols and their evolution in the process of aggregation: (1)
initial aggregation stage; (2-4) hydrosol prepared on the basis of
collargol at various stages of aggregation; and (5) hydrosol
prepared by the procedure \cite{9,39,43} using NaBH$_4$. Curves 4
and 5 correspond to the developed stage of hydrosol aggregation.}
\end{center}
\end{figure*}

\begin{figure*}[!h]
\begin{center}
\includegraphics[height=.5\textwidth]{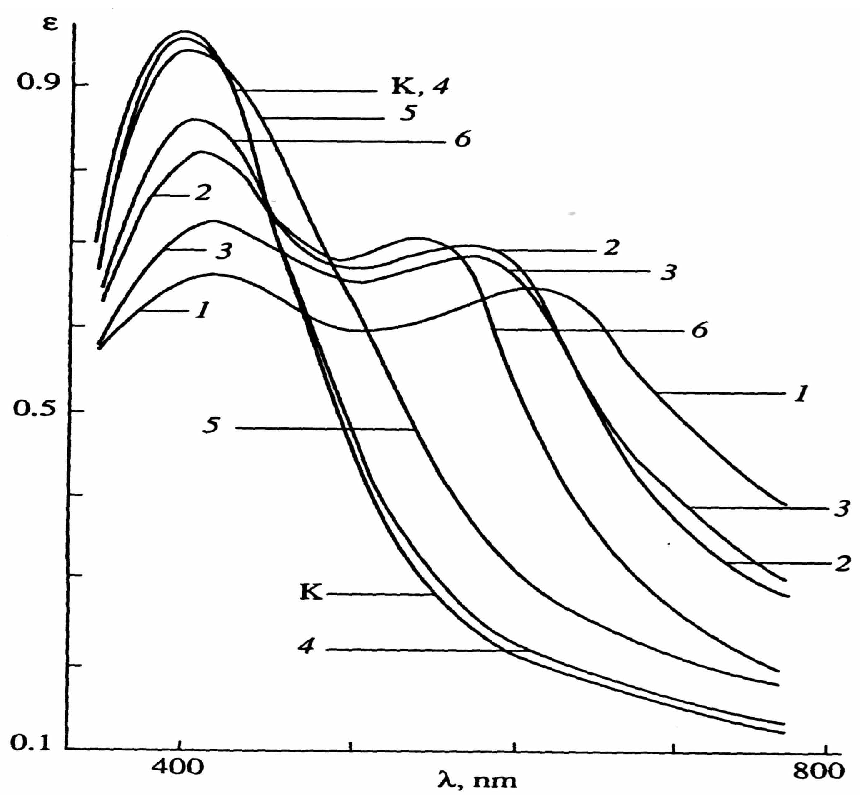}
\caption[recon] {\label{f5}  Changes in the absorption spectra of
Ag C-hydrosol during aggregation. The colors of the solutions are:
(K) yellow; (1) dense purple blue; (2) dense brown; (3) brownish
red; (4) yellow; (5) dark orange; (6) red.}
\end{center}
\end{figure*}

Silver sols can be considered as a convenient model system where the effect
of various factors on the position and shape of each single resonance can be
easily monitored, thanks to the presence of the isolated and
relatively-narrow band of plasmon absorption centered at $\lambda_{pl}$ in
the optical range.  Figure 4 illustrates the typical absorption spectra of
various silver hydrosols,  which comprise  both isolated spherical particles
(the average size of the metallic core is  $2R_i = 14 -16$ nm) and particles
combined into fractal aggregates. Procedures for the preparation of
hydrosols are described above (see also \cite{1,9}).  A typical feature of
the spectral curves of the majority of strongly-aggregated
surfactant/polymer-containing silver colloids (see, e.g., Fig. 4, curve 4
and  Fig. 5, curve 1) is the presence of two maxima in the optical spectral
range \cite{1,9,42}. The first spectral maximum corresponds to the surface
plasmon of isolated and weakly-interacting particles when their number is
the largest. Its amplitude reduces with an increase in the degree of sol
aggregation due to a decrease in the relative fraction of such particles
(Fig. 4, curves 2 - 4). We believe that the origin of the second maximum
fundamentally differs from that of the first maximum. A possible explanation
of the origin of the second low-energy peak in the absorption spectra of
silver sols is reviewed in \cite{10}. This problem was also touched upon in
\cite{9,43}, where it was suggested that the second maximum is associated
with the Raman excitation band. The origin of this maximum, explained by the
excitation of the collective transverse waves in the ensemble of particles,
has been studied within the framework of the Maxwell-Garnett theory
\cite{44}. However, this viewpoint contradicts the fact that fractal
aggregates forming in a colloid cannot transmit travelling waves due to the
violation of the condition of translational invariance in fractal
structures.

In general, we believe that the explanations represented in basic
publications devoted to this problem are often confined to the framework of
qualitative hypotheses, and the proposed theoretical models in many cases
contradict experimental facts. Meanwhile, the appearance of an additional
low-frequency maximum in the absorption spectrum of an aggregating sol and
the gradual shift of this maximum within the background of forming a
long-wavelength wing give rise to sharp changes in color; this situation is
discussed in detail in \cite{1}. We also note that the appearance of a
low-energy maximum in the absorption spectra in some cases is observed not
only in hydrosols, but also in silver aerosols deposited onto the substrate
\cite{10}.

\subsubsection{Evolution of the spectra of silver hydrosols during
photostimulated aggregation of the disperse phase}

In darkness, the colloid looses it stability very slowly (over a period
ranging from a few weeks to a few months) due to random Brownian collisions
of particles and low coagulation efficiency of such collisions. However, the
aggregation stability of the colloid drops abruptly under the action of UV
or visible radiation. As reported in \cite{41}, we observed photoinduced
formation of fractal aggregates under the action of the radiation of certain
lasers or nonmonochromatic (or quasimonochromatic) radiation on hydrosols.
This effect is also accompanied by considerable broadening of absorption
spectra. The difference between the absorption spectra of spontaneously
formed and photostimulated aggregate is negligible at low light intensities
\cite{1} (cf. Fig. 5 and Fig. 6).

\begin{figure*}[!h]
\begin{center}
\includegraphics[height=.5\textwidth]{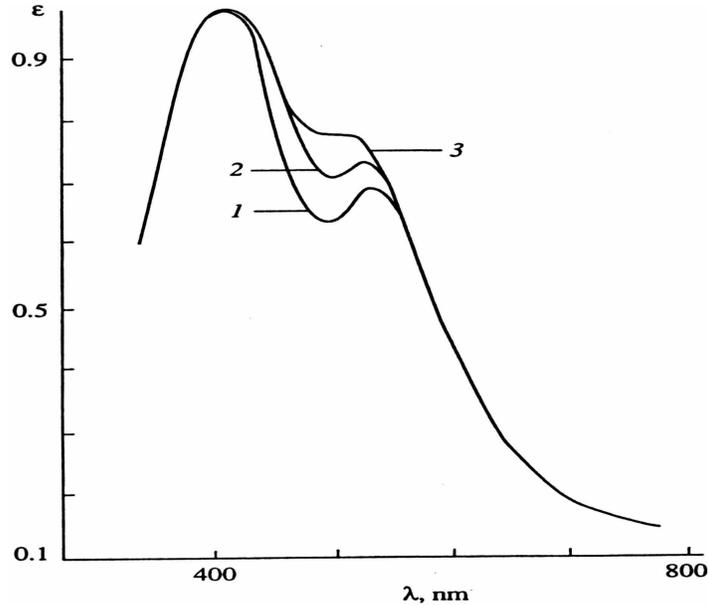}
\caption[recon] {\label{f6} Changes in the absorption spectrum of
Ag A-hydrosol occurring in darkness after preliminary exposure to
light with the wavelength $\lambda$ = 500 nm, intensity $I$ = 3
mW/cm$^2$ and exposure time of 10 hours. The time between the
changes corresponding to curves 1 and 2 was 6 days. The time
between the changes corresponding to curves 2 and 3 was 8 days.
The colors of the solutions are: (1) red; (2) dark red; and (3)
greenish brown. }
\end{center}
\end{figure*}

It appears only under irradiation by the pulsed lasers and grows with the
increase of the intensity of the stimulating light (Fig. 7).

\begin{figure*}[!h]
\begin{center}
\includegraphics[height=.5\textwidth]{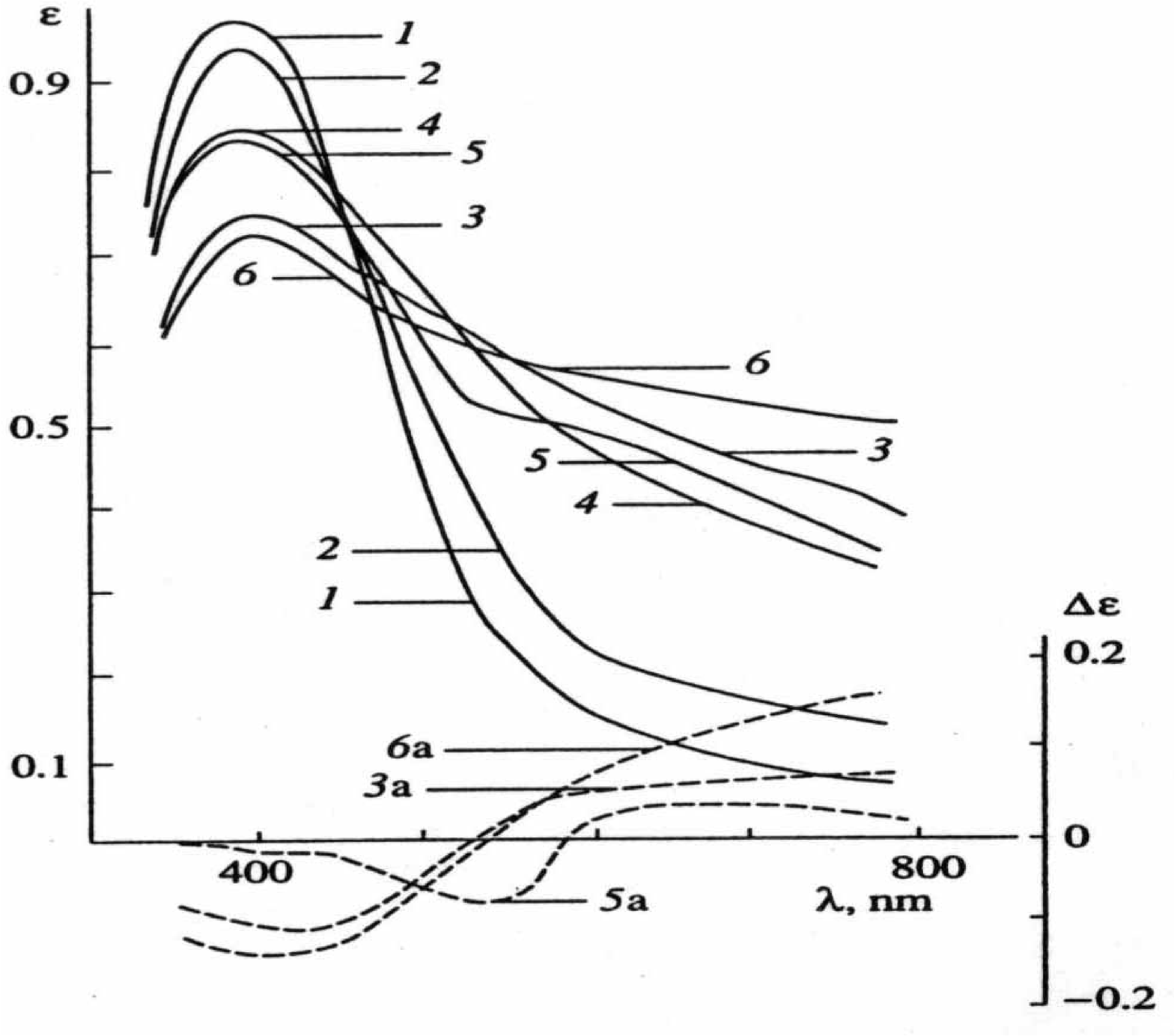}
\caption[recon] {\label{f7} Changes in the absorption spectra of a
Ag hydrosol caused by laser radiation: (1) initial spectrum of a
non-aggregated hydrosol; (2) changes in the spectrum after
exposure of the solution to the radiation of an argon laser
($\lambda$ = 514.5 nm, intensity 50 mW/cm$^2$, exposure time of 2
hours); (3) changes in the spectrum after irradiation of the
solution with 15 pulses of a ruby laser ($\lambda$ = 694 nm, $W$ =
3 J/cm$^2$, $\tau$ = 30 ns); (4) initial spectrum of a Ag hydrosol
before its exposure to the radiation of an YAlO$_3$ : Nd laser;
(5) changes in the spectrum after irradiation with 20 pulses of
the second harmonic of an YAlO$_3$: Nd laser ($\lambda$ = 540 nm,
$W$= 1.5 mJ/cm$^2$, $\tau$ = 30 ps); and (6) changes in the
spectrum after irradiation with 250 pulses of an excimer (XeCl)
laser ($\lambda$ = 308 nm, $W$ = 90 mJ/cm$^2$, $\tau$ = 30 ns);
(3a, 5a, 6a) differential spectra of the hydrosols (with respect
to curve 4). The values of $\varepsilon^{max}(\lambda_{pl})$ for
curves 3, 5, and 6 are different because of different radiation
doses (for explanations, see the text).}
\end{center}
\end{figure*}

Figure 7 depicts the absorption spectra of silver hydrosols modified by
laser radiation. The hydrosols were exposed to radiation continuously
scanning the solution surface of 1 $\times$ 4 cm in a quartz cuvette (cell)
0.2-cm thick. Curve 2 demonstrates the changes in the spectrum that occurred
upon irradiation by an argon laser. The spectrum shows signs of the starting
stage of aggregation. Curve 3 shows  changes in the absorption spectrum that
occurred after exposure to the pulsed radiation of a ruby laser ($\lambda$ =
694 nm). Apart from the changes representing the general trend (such as the
formation of a long-wavelength wing), the spectrum also displays a feature
associated with the considerable decrease of absorption in the area of 694
nm (in the range $550 - 750$ nm). This decrease is clearly visible in the
difference spectrum (curve 5a). The reason for such a decrease in absorption
is that the formation of fractal aggregates (FAs) and the occurrence of an
inhomogeneous wing of the spectrum create conditions for the subsequent
spectrally-selective photodestruction of the FAs (near the wavelength of the
laser radiation). We note that the photomodification of FAs takes place only
when the radiation energy density  exceeds the threshold value. A similar
trend is observed in hydrosols exposed to radiation of the second harmonic
of a neodymium laser ($\lambda$ = 540 nm). Curve 4 represents the spectrum
of the original hydrosol in the intermediate stage of aggregation (unlike
curve 1). Curve 5 demonstrates the changes in the absorption spectrum that
occurred after irradiation with the laser pulses. Aside from the formation
of a long-wavelength wing, the spectrum seems to develop a dip in the
vicinity of the wavelength of the laser radiation (540 nm). Such spectral
consequence of two processes -- photoaggregation and photomodification. We
note that a considerable increase of the threshold radiation dose results in
deterioration of the spectral selectivity of irradiation and in a wider dip,
as can be seen, for example, in curve 3. Special attention  should be paid
to curve 6, which demonstrates spectral changes with signs of aggregation.
These changes are brought about by the action of the pulsed radiation of an
excimer (XeCl) laser on the colloid. Curve 6 in Fig. 7 does not have a
second maximum unlike curve 6 in Fig. 5, which represents the spectral
changes caused by low-intensity radiation. This difference is most probably
associated with the photomodification of FAs. The hole (dip) burnt in the
absorption spectral range $\lambda>\lambda_{pl}$, while
$\lambda_{las}>\lambda_{pl}$, gives rise to a corresponding dip in the range
$\lambda<\lambda_{pl}$, and vice versa in the case where
$\lambda_{las}<\lambda_{pl}$. According to the theory in \cite{13}, the
short-wavelength spectral dip is twice as far from $\omega_{pl}$ (on the
frequency scale), compared to that in the long-wavelength wing of the
absorption spectrum.  This fact, along with the relationship between
$\lambda_{las}$, and $\lambda_{pl}$, brings us to the conclusion that the
hole burnt by the eximer laser in the range $\lambda>\lambda_{pl}$ falls
within spectral range of the long-wavelength absorption maximum (second
maximum on curve 6).

It should be noted that exposure of the solution to the radiation of a
neodymium laser with the wavelength $\lambda$ = 1.08 $\mu$m causes no
spectral changes. This fact can also be explained within the suggested model
of the phenomenon, because the wavelength of this laser radiation lies
beyond the red threshold of photoaggregation, which is believed to be of a
photoemissive nature \cite{37,38}.

\subsection{Main factors determining optics of sol in
the theory of OPFC} In this work, specific features of the absorption
spectra of silver sols and their differences within the framework of the
theory of OPFC are explained based on a concept which elaborates the
approach proposed in \cite{1}. The essence of this idea is as follows. In
general, any real sol is a polydisperse system characterized by the function
of particle size distribution (FPSD) $f(2R_i$), which has a
clearly-pronounced maximum and is described by the asymmetric dependence
resembling a Poisson distribution. This means that the most probable
particle size $2R_m$ exists in  a system that has real colloidal properties
(for example, see \cite{9}). In this sense, the monodisperse system can be
considered as a specific (with the narrowest FPSD) system, although this
case is physically meaningless.

Correspondingly, the presence of a prevailing, statistically-sampled size
$2R_m$ leads to the fact that, over the course of aggregate growth, the most
probable distances between contacting particles become those which arise in
the pairs with particle size $2R_m$ (see the explanation in Sec. 4.4).  In
turn, the appearance of an excess number of contacting pairs with the most
probable interparticle distance should affect the optical spectra. This is
related to the fact that these pairs correspond, on average, to the
prevailing value of the frequency resonant shifts  $(\Delta\omega_m)_i$; the
latter circumstance is responsible for an increase in absorption over some
limited part of the long-wavelength wing of an aggregate spectrum.  Within
the framework of the pair approximation (with no allowance for the exact
total contribution of distant particles), the relative position of the
second maximum $\omega_2$ in the absorption spectrum is determined by the
relations
\begin{equation}\label{4.1}
\omega_2\propto\omega_r-(\Delta\omega_m)_i;\quad (\Delta\omega_m)_i\propto
(2R_m)^{-3}_{ij},
\end{equation}
where $\omega_r$ is the resonant frequency of noninteracting particles.
However, in this case, it should be kept in mind that, in
surfactant/polymer-containing sols, the least possible distances between the
particle geometrical centers $2(R_m)_{ij}$ are determined both by the size
of the particle metallic core and the thickness $L$ of their adsorption
layers as a whole, including the polymer component, which provides for the
appearance of a sterical factor of stability \cite{45}. As is seen from
microimages of fractal aggregates of different-type silver hydrosols
\cite{42}, the polymer component of the adsorption layer markedly affects
the packing of particles in aggregates. In the case of identical $L$ values
for all the particles, the probability of the appearance of the pairs of
contacting particles with interparticle distance $2(R_{m0})_{ij}=2(R_m
+L)_{ij}$ (with allowance for the deformation of the external part of their
adsorption layers) in the optical spectrum of the aggregated hydrosol should
be the largest.  In addition, the case should also be considered when the
thickness of adsorption layers on particles of various sizes can be
different. Under certain conditions, an $L(R_i)$ dependence can appear.
Moreover, taking into account differences in the composition of the
dispersion medium and the type of stabilizer molecules, various types of
colloids can be characterized by the strictly individual pattern of this
function. As will be shown further, this can be one of the very reasons for
the differences in the adsorption spectra of some silver hydrosols (with
account for differences in the FPSD for the metallic core) represented, for
example, in  \cite{1,42}.

The reason for the differences in the adsorptivity of particles (including
the adsorptiviity with respect to surfactant or polymer molecules) can be
related, in particular, to the presence of the defects (vacancies) of a
crystalline lattice at the real particle surface. This leads to the
appearance of local regions with a nonequilibrium value of the electric
potential at the particle surface and, consequently, to the selectivity of
various parts of the surface with respect to the electrostatic interaction
with molecules of the adsorption layer. Therefore, in the process of
adsorption, only those parts of a particle surface that are located near the
vacancies being the adsorption sites are first occupied \cite{46,47}. The
adsorptivity of a particle as a whole will depend on the surface density of
vacancies, including those that arise in the process of self-induced
adsorption \cite{46}. As was demonstrated in this work, the number of such
vacancies can surpass the number of equilibrium surface vacancies by many
orders of magnitude. The dependence of vacancy concentration on the particle
size $C_v(R_i)$, which rises exponentially with a decrease in $R_i$, was
revealed in \cite{46} for small particles.  Given what has been said above,
the most probable distances between pairs of particles in aggregates will
generally correspond to the maximum of FPSD, $F^{max}(R_i)=F(R_{m0})$, where
$F(R_i )=f(R_i )L(R_i )$. It is this function that will affect the most
probable values of the frequency shifts which, within the framework of the
binary model with  allowance for known constraints of this approximation,
are described by the expression
\begin{equation}\label{4.2}(\Delta\omega_{m0})_i\propto(2R_{m0})_{ij}^{-3}.
\end{equation}
However, the problem of the effect of the surfactant/polymer-component of
the adsorption layer on the pattern of particle packing in fractal
aggregates, and hence on the spectral features of these aggregates, can be
solved by experimental studies of the thickness of adsorption layers using
electron microscopy. In this case, a number of complexities related to the
preparation of microscopic samples can arise. This will require special
techniques, because the dehydration of surfactant/polymer molecules can be
accompanied by a variation in molecular volumes and, correspondingly, by a
decrease in the initial thickness of an adsorption layer in the vacuum
chamber of an electron microscope.

Hence, in accordance with the above discussion, one can state that, in the
formation of the contour of the long-wavelength wing of the absorption
spectrum, two factors are manifested: (1) on the one hand, spatial
disordering and local anisotropy of fractal aggregates, which takes place in
any type of sol, irrespective of the pattern of the distribution function,
and generally specifies a monotonic decay of absorption with increasing in
wavelength; and, (2) on the other hand, the existence of a dominating
particle size in a real polydisperse colloid. The introduction of FPSD with
a selected maximum into the equations of the theory of OPFC leads to a
violation of the monotonicity in the boundaries of the long-wavelength wing
of the absorption spectrum of a fractal aggregate. To solve this problem, it
is necessary only to determine the conditions which, in a certain size
range, can result (via the mediated effect on the aggregate structure) in
the appearance of an additional low-frequency maximum in the absorption
spectrum of a fractal aggregate, provided that the dominant particle size in
some types of metal sols is taken into account.

\subsection{ Analysis of calculated spectral dependencies}

While calculating the absorption spectra of fractal aggregates, we took into
account the results of \cite{32}, where important refinement was introduced
into the theory of OPFC for the calculation of linear optical spectra. This
refinement is related to the allowance for the real particle sizes, and
accordingly, to the anisotropic fields of oscillating dipoles induced on
interacting particles. According to this work, the value of spectral
broadening observed in real fractal aggregates should correspond to shorter
interparticle distances than the least possible distances confined to the
condition of sphere contact. This implies that, to adequately describe the
spectrum of real fractal aggregate  consisting of contacting particles, it
is necessary to assume that the particle size used in the calculations
should be smaller than the experimental size by the correction coefficient
whose average value is equal to $K\approx 1.65$ (by the data of various
authors). In this case, the account for this correction is nothing other
than the procedure of renormalization drawing together calculated spectra
with a given value of particle sizes and experimental dependencies.

\begin{figure*}[!h]
\begin{center}
\includegraphics[height=.6\textwidth]{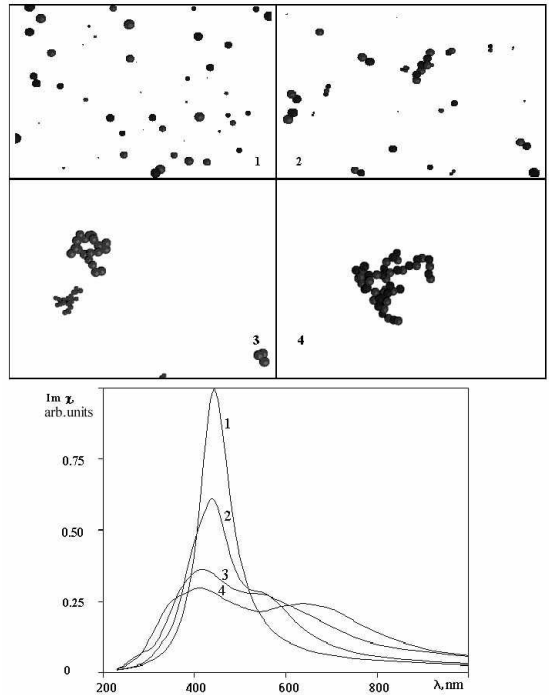}
\caption[recon] {\label{f8} Simulation of 3-D coagulation in the
ensemble of 50 monodisperse silver particles with a diameter of 20
nm (upper) and the calculated corresponding evolution of spectral
dependencies of the imaginary part of their linear optical
susceptibility (absorption spectra)(lower). }
\end{center}
\end{figure*}

Figure 8 illustrates the results of calculations of the wavelength
dependence of the imaginary part of the linear optical susceptibility of an
aggregating particle ensemble corresponding to its absorption spectrum (for
the case of a monodisperse sol). Hereafter, we adhere to the terminology
introduced in \cite{13} where $\Im\chi(\omega)$ appears as an absorption
spectrum. (The absorption cross-section differs  from this value only by a
trivial factor of $4\pi k$.) Figure 8 shows the regularities of the spectrum
evolution at various stages of particle aggregation (curves 1-4). The
formation of the second spectral maximum is already observed at the
intermediate stages of aggregation. When comparing this set of curves with
the experimental results (Fig. 4), attention should be given to the obvious
qualitative similarity of these dependencies.

\begin{figure*}[!h]
\begin{center}
\includegraphics[height=.25\textwidth]{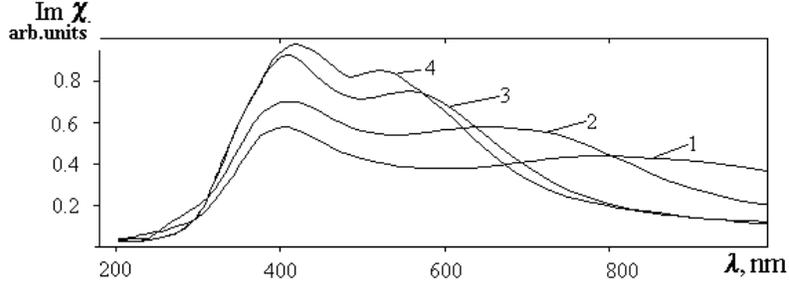}
\caption[recon] {\label{f9} Variations in the absorption spectra
of fractal aggregates formed from monodisperse particles as a
function of their sizes ($2R_m$): (1) 18.2, (2) 19.8, (3) 21.6,
(4) 23.1 nm.}
\end{center}
\end{figure*}

Figure 9 represents the absorption spectra of fractal aggregates
formed from particles of identical sizes. In the obtained set of
curves, the spectrum pattern is studied as a function of particle
size. These sizes correspond to the following minimal values of
distances between the nearest particles: $R_{ij} = 11, 12, 13, 14$
nm. With allowance for the coefficient $K$, this corresponds to
the particle sizes of $2R_m = 18.2,  20,  21.6, 23.1$ nm. As seen
from the figures, in accordance with the expressions (\ref{4.1})
and (\ref{4.2}), clear correlation between the value of $2R_m$,
and the position of the long-wavelength spectral maximum
$\omega_2$ is observed. Hence, the smaller the particle size, the
stronger the secondary maximum is shifted. A gradual decrease in
its contrast is also observed. The following important fact should
be mentioned: if the dominant particle size in an aggregate
becomes smaller than $15-18$ nm, the secondary maximum completely
shifts beyond the boundary of the optical range and has no further
significant influence on colloid color. If the characteristic
particle size exceeds $25 - 30$ nm, the position of a secondary
maximum appears to be too close to the principal maximum
($\lambda_r$) and gradually vanishes into the background of the
long-wavelength wing of the principal spectral maximum with an
increase in particle size. In this case, color changes can only be
attributed, as in the case of $2R_m < 15 - 18$ nm, to a monotonic
rise in the absorption in the region of the long-wavelength wing.
Note, however, that results shown in Fig. 9 are only valid for
ideal monodisperse colloids, which is physically an unreal case.
However, this does not mean that particles with the sizes larger
than 30 nm do not contribute to the spectrum broadening. A
contribution can also be made by larger particles if they comprise
interacting pairs including small particles.  However, the
distance between the particle centers of such a pair should fall
within the range of $R_{ij}$ shown in Fig. 1 (with account for the
factor  $K$), although it is evident that the number of such
pairs,  and hence their spectral contribution, is insignificant
compared to the particles of dominant size. Thus, the
fundamentally important result of these calculations is the fact
that the revealed specific features of absorption spectra of
aggregated colloids can only be exhibited within a rather narrow
range of particle sizes.

\begin{figure*}[!h]
\begin{center}
\includegraphics[height=.5\textwidth]{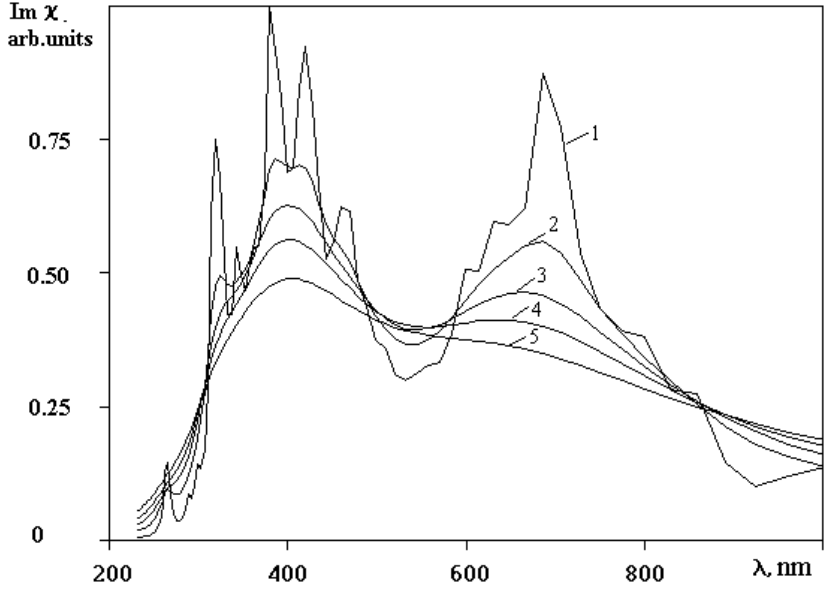}
\caption[recon] {\label{f10} Absorption spectra of silver fractal
aggregates at various values of the homogeneous width of
absorption spectrum ($\Delta\lambda$) attributed to the particles
comprising an aggregate:  (1) 50, (2) 80, (3) 100, (4) 120, and
(5) 150 nm.}
\end{center}
\end{figure*}

Figure 10 shows the results of the calculations of absorption
spectra of aggregates formed for various values of the homogeneous
width ($\Delta\lambda$) of the absorption spectrum of comprising
particles. These results solve the problem of why there is a
decrease  in the contrast of the secondary maximum in the
spectrum. As is seen from these dependencies, statistical
oscillations appear in the spectrum of fractal aggregates at
$\Delta\lambda$ values of up to 50 nm (below minimal values
observed in the experiment). These oscillations disappear at
$\Delta\lambda = 90 - 100$ nm. Values of the spectral width of the
surface plasmon of silver hydrosols with zero-degree aggregation
\cite{37} not larger than 90 nm were observed in \cite{42}.
However, the $\Delta\lambda$ values can be slightly larger in some
hydrosols. This can be related, firstly, to spectral broadening
due to the presence of a some fraction of small particles with
large values of homogeneous width (see formula (\ref{1.1})) and,
secondly, to the presence of microscopic aggregates composed of
several particles that already exhibit the effect of spectral
broadening resulting from their interaction. In addition, a
certain polydispersity of the sol, even in the range of small
particles, can be one of the reasons for the presence of a slight
dispersion of the "resonant" frequency $\omega_r$ \cite{5,6} and,
hence, of some inhomogeneous spectral broadening due to the
existence of the weak $\omega_r(R_i)$ dependence in this range.
Finally, it is necessary to take also into account some
nonsphericity of a small particle fraction, which can result in
the shift of resonances depending on the particle shape
\cite{5,6}. Note that the presence of oscillations in the obtained
spectral dependences at $\Delta\lambda = 80$ nm, varying in the
range of experimental values of the homogeneous width of an
absorption spectrum for some types of silver colloids \cite{42},
can also be explained by the restriction imposed in the
calculations for the number of particles being 50 in a fractal
aggregate. As a result, the number of statistical averagings seems
to be insufficient for calculating the spectral curve with a
narrow homogeneous bandwidth.

\begin{figure*}[!h]
\begin{center}
\includegraphics[height=.4\textwidth]{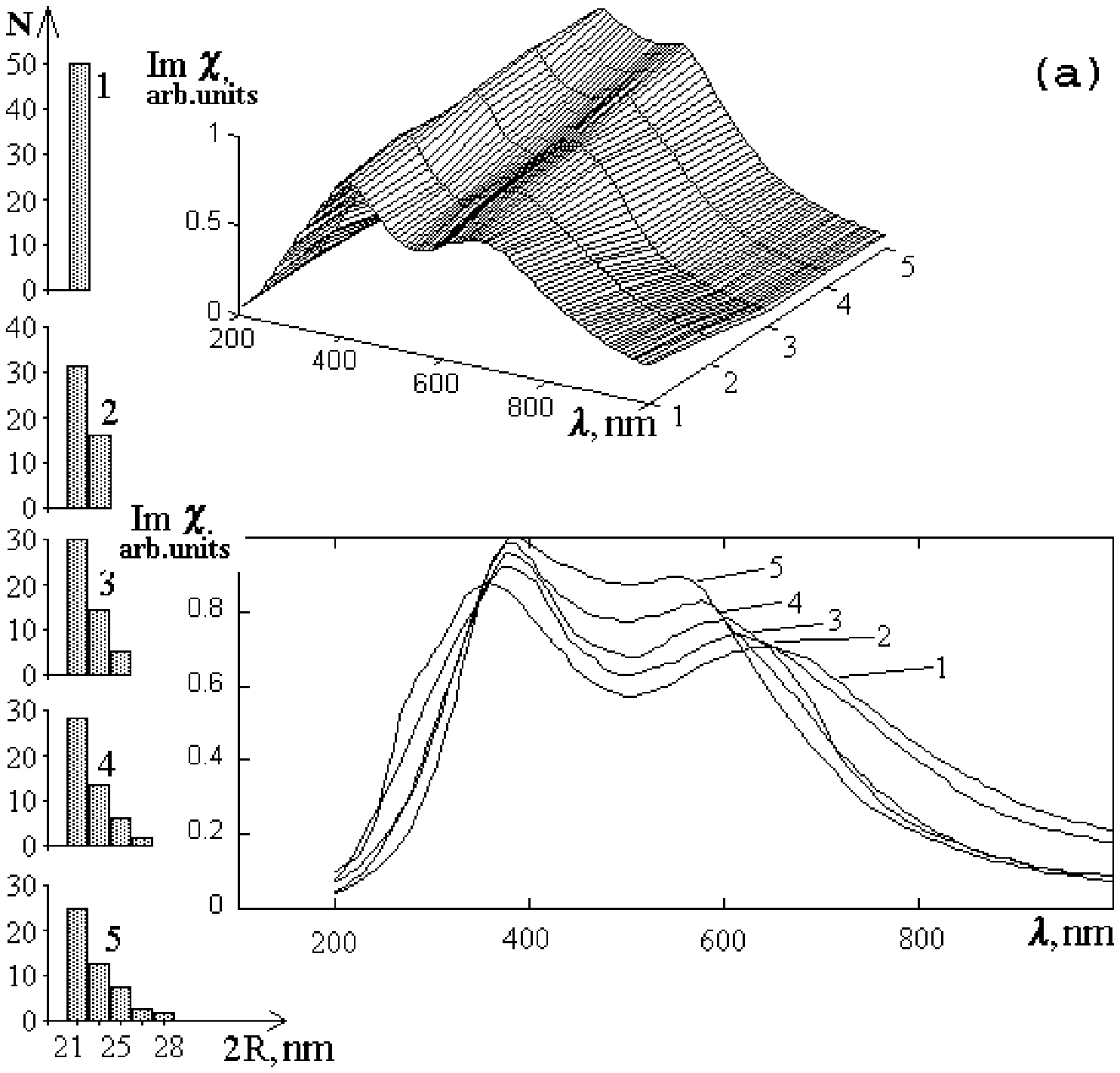}
\includegraphics[height=.38\textwidth]{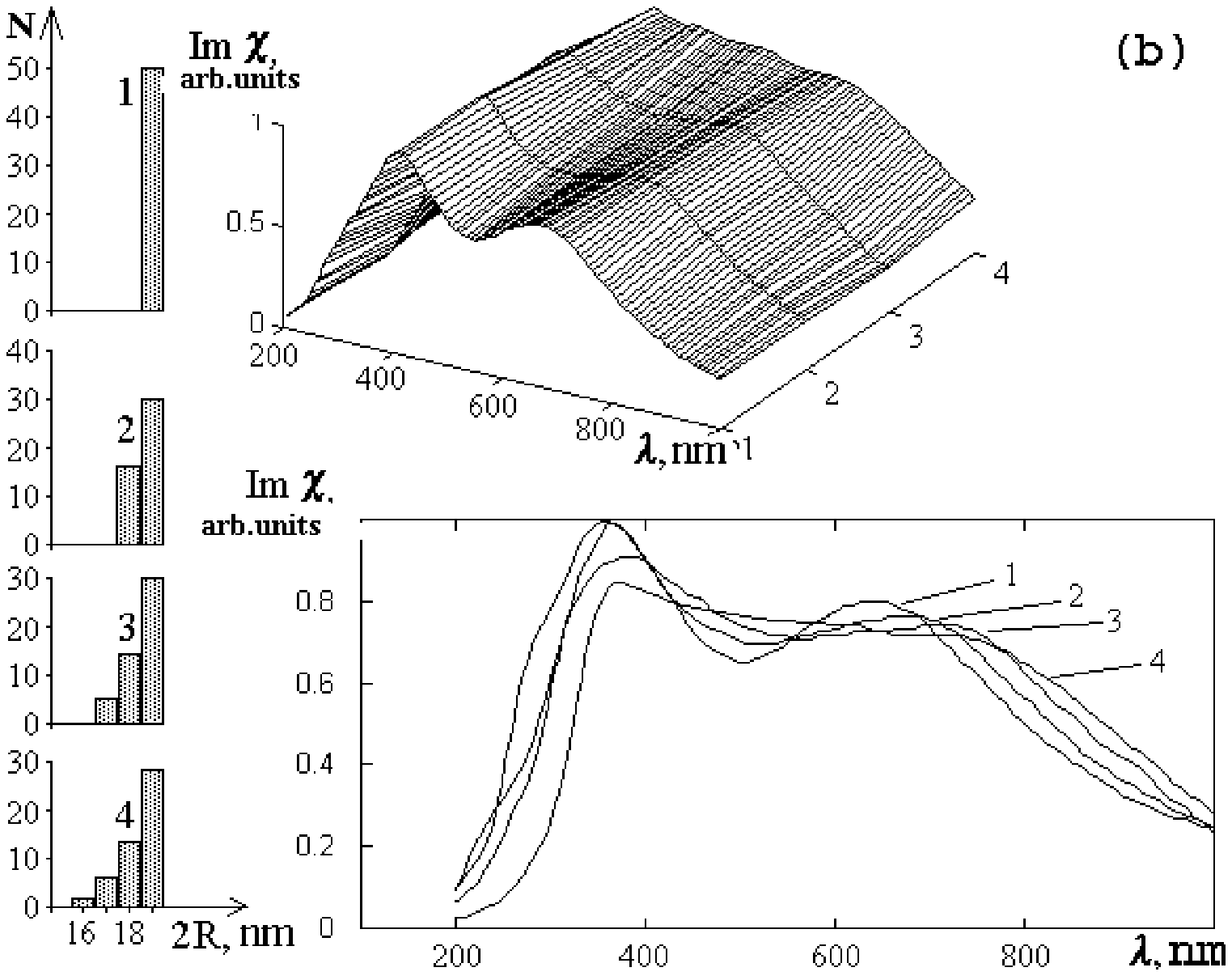}\\
\includegraphics[height=.4\textwidth]{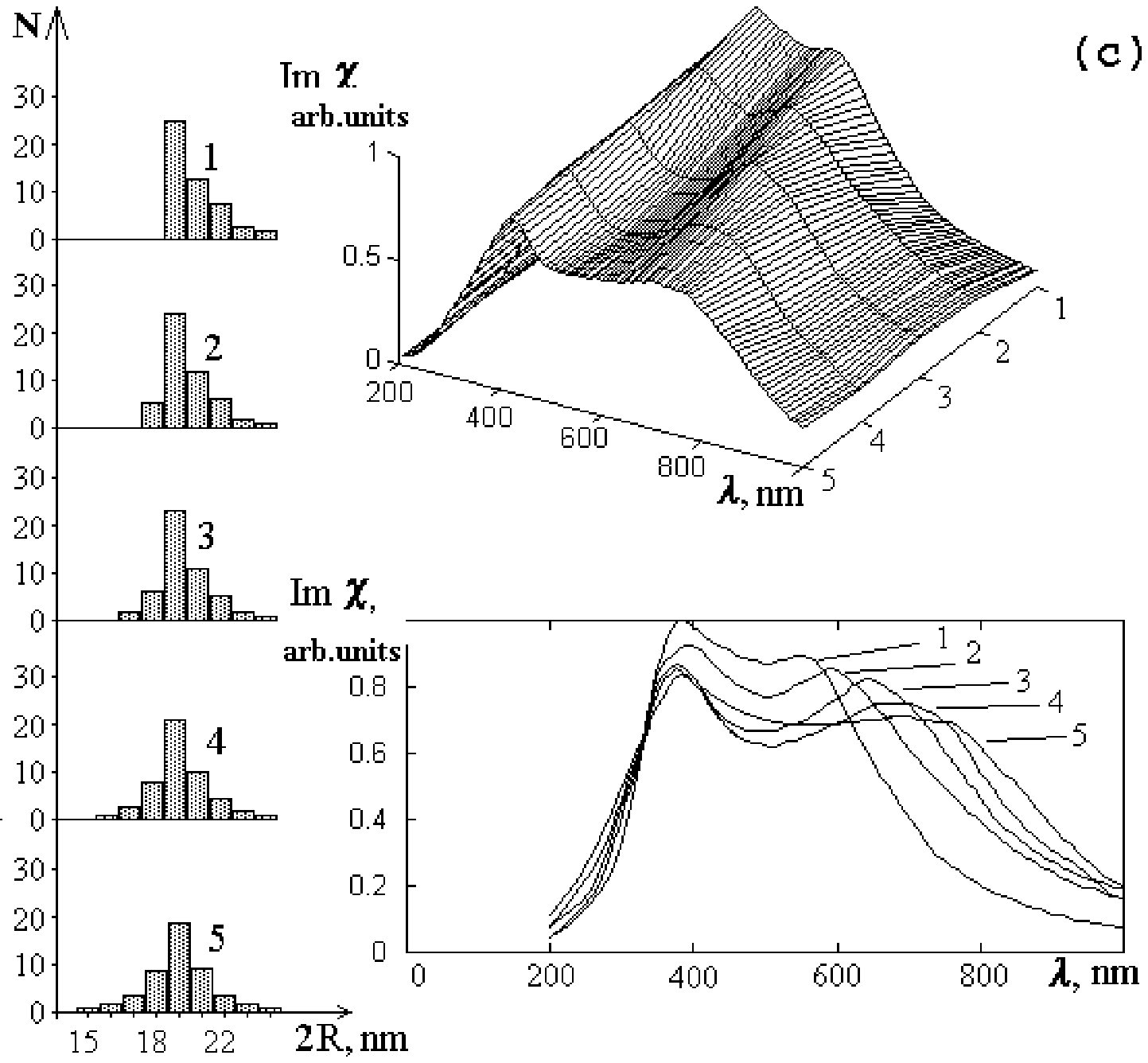}
\caption[recon] {\label{f11} Absorption spectra of fractal
aggregates of a polydisperse ensemble of particles with the
homogeneous absorption spectrum of an isolated single particle
$\Delta\lambda = 100$ nm but for different distributions of the
particles over their sizes. (a) The variation in the distribution
is performed by adding larger particles to  particles of the same
size of $2R_i$ = 21 nm ($N$ is the number of particles). (b) A
polydisperse sol is prepared by adding smaller particles to a
monodisperse system with $2R_i$ = 21 nm. (c) A polydisperse sol is
prepared by adding smaller particles to a polydisperse sol
containing larger particles, corresponding to the statistical
distribution in the initial polidisperse sol in the case 5
depicted  in part (a) above. The plot (c) displays the
conservation of the contrast of the secondary maximum in a
spectrum, which is combined with a significant shift of this
maximum to the long-wavelength range.}
\end{center}
\end{figure*}

Studies of the absorption spectra of polydisperse ensembles of particles
with the FPSD close to a real distribution \cite{9} are of the most
practical interest in this work. Figure 11a demonstrates absorption spectra
of a polydisperse ensemble of particles ($\Delta\lambda = 100$ nm) with
various types of FPSD. As was demonstrated in preliminary calculations, the
spectral dependence turned out to be extremely sensitive to the form of
FPSD. Therefore, we passed from the monodisperse to polydisperse cases in
two stages. At the first stage, to the particles of a specified size ($2R_m
=21$ nm) whose aggregate spectral position of secondary maximum corresponds
to the experimental data ($\lambda_2 = 600 - 620$ nm), larger particles were
gradually added, which eventually provides for the similarity of this wing
of FPSD with the $F(2R_i )$ \cite{9} of a real sol. As seen from the set of
curves obtained, the common general feature is the shift of the secondary
maximum to the principal one upon the addition of larger particles to the
system.

However, the largest sensitivity of the absorption spectra to the variation
in FPSD while passing to polydisperse systems is observed for the appearance
of small particles in a system. Figure 11b represents the set of curves
obtained upon the gradual addition of smaller particles to a monodisperse
system ($2R_m = 21$ nm), providing for the monotonic decay of the FPSD wing.
As seen from these results, the regularity is exhibited in the shift of the
secondary maximum to the long-wavelength range and in a decrease in its
contrast. The latter is due to the broadening of the low-frequency spectral
peak (Fig. 1) with a decrease in the distance between the pair of
interacting particles brought into contact. In turn, a decrease in the
interparticle distance is explained by a corresponding decrease in particle
sizes.

Finally, Fig. 11c demonstrates the set of curves obtained upon adding small
particles to a polydisperse sol containing only large particles (FPSD
corresponds to case 5 in Fig. 11a when the condition of the preservation of
the contrast of the secondary maximum in a spectrum is accompanied by the
largest shift of a maximum to the long-wavelength range). As seen from Fig.
11c, the gradual addition of small particles to the system leads to the
shift of the secondary maximum to the long-wavelength range and
simultaneously to a decrease in its contrast. We see that curve 4 best
corresponds to the experimental results. Note the asymmetric pattern of FPSD
corresponding to some deficiency of small particles, which also correlates
with the experimental data \cite{9}. In the spectrum of the fractal
aggregate obtained for asymmetric FPSD (Fig. 11c, case 5), the contrast of
the secondary maximum falls sharply. A common tendency is that the
appearance of an excess number of small particles in an ensemble results in
the gradual disappearance of the secondary maximum in the spectrum; in this
case, the long-wavelength wing of the absorption spectrum of such colloids
is described by a smooth monotonic dependence.

Hence, results of these calculations are important, because the
long-wavelength maximum can be most clearly pronounced at the least possible
width of FPSD. Evidently, the greater the FPSD width, especially with an
excess of small particles, the more blurred the secondary maximum becomes.
This feature can explain the absence of the secondary long-wavelength
maximum in the absorption spectra of silver hydrosols prepared by reducing
silver nitrate with NaBH$_4$ (Fig. 4, curve 5) \cite{9,43}. The absence of a
surfactant/polymer component in the adsorption layer composed of particles
of these silver colloids stabilized by electrostatic interactions results in
the average effective particle size and, correspondingly, average
interparticle distances in aggregates becoming smaller than in the
aggregates of collargol-based colloidal silver \cite{1,42}, even if the
FPSDs of these colloids with respect to metallic core are similar.

\begin{figure*}[!h]
\begin{center}
\includegraphics[width=0.5\textwidth]{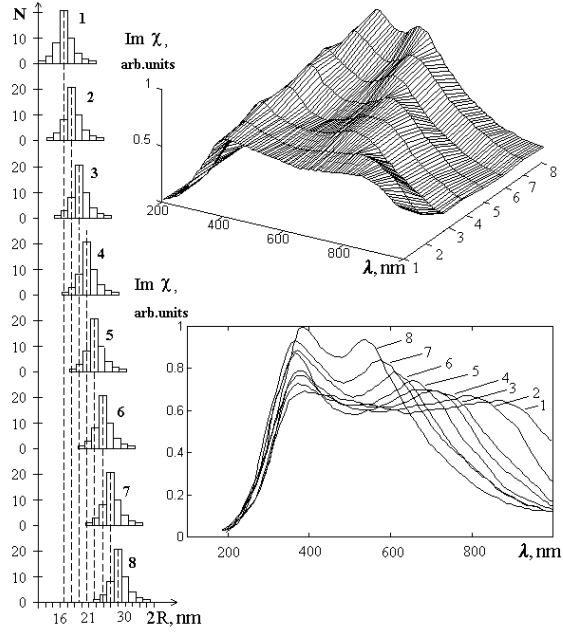}
\caption[recon] {\label{f12} Dependence of the absorption spectra
of fractal aggregates of polydisperse silver sols, having
identical size-distribution profiles, on the position of the
distribution maximum, $2R_m$: (1) 16.5, (2) 18.2, (3) 19.8, (4)
21.5, (5) 23.1, (6) 24.8, (7) 26.4, (8) 28 nm.}
\end{center}
\end{figure*}

Figure 12 shows calculated dependences of absorption spectra of polydisperse
fractal aggregates (with identical FPSD profiles) on the position of the
maximum of this function. As seen from this figure, the common tendency is
manifested in the shift of the secondary maximum into the long-wavelength
range upon the displacement of the FPSD maximum towards smaller sizes, which
is also attributed to the prevailing contribution of small particles. It is
of some interest to compare these results with the data shown in Fig. 9 for
a similar dependence for monodisperse sols. As seen from comparing these
figures, the colloid polydispersity results in that spectral dependences,
whose secondary maximum positions are similar, are already observed at the
values of the FPSD maxima that slightly exceed the corresponding sizes of
particles comprising monodisperse fractal aggregates.

\begin{figure*}[!h]
\begin{center}
\includegraphics[width=0.5\textwidth]{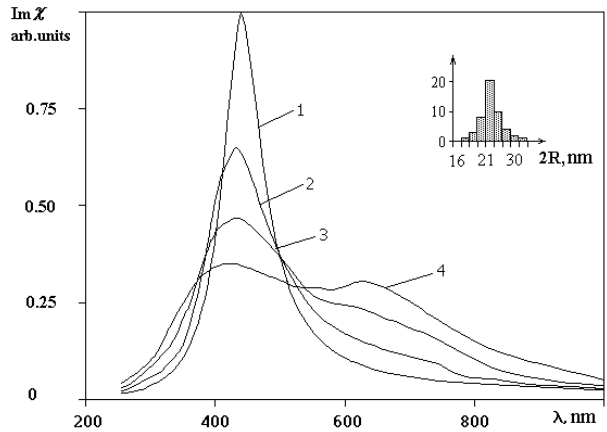}
\caption{\label{f13}  Absorption spectrum of a polydisperse
ensemble of silver particles with maximum size-distribution at
$R_m = 23.1$ nm (corresponding to the case 5 in Fig. 12). Curves 1
- 4 display the spectrum evolution at various stages of the
aggregation process (cf. Figs. 4 and 8).}
\end{center}
\end{figure*}

Figure 13 demonstrates the evolution of an absorption spectrum of a
polydisperse ensemble of particles with the FPSD pattern maximally close to
the real one ($2R_m =23.1$ nm). A comparison with Fig. 10 representing
analogous dependences, however for a monodisperse ensemble of particles
($2R_m =20$ nm), reveals their qualitative similarity.

In this work, we attempted to find out why the principal maximum
in the absorption spectra of some strongly-aggregated real
colloids is close (or almost coincides) to this maximum (Fig. 4),
whereas the results of calculations indicate a certain shift
towards a long-wavelength range. It was shown in this work, one of
the reasons for this situation can be the presence of some amount
of isolated particles, which are not included within fractal
aggregates.
\begin{figure*}[!h]
\begin{center}
\includegraphics[width=0.5\textwidth]{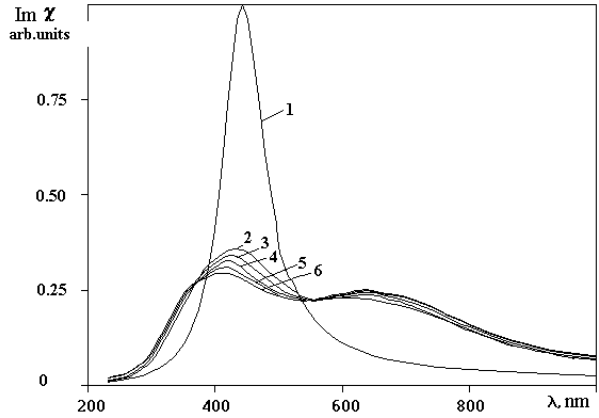}
\caption[recon] {\label{f14} Variation in the absorption spectra
of an ensemble containing fractal aggregates and an additional
small fraction $f$ of free isolated particles as a function of the
value of this fraction: (1) spectrum of isolated particles. (2 -
6) spectra of the fractal aggregate composed of 50 particles in
the presence of free particles at $f$= 1/10, 1/12.5, 1/17, 1/25,
1/50, respectively.}
\end{center}
\end{figure*}

Figure 14 demonstrates the variation in absorption spectra of a particle
ensemble containing fractal aggregates and an additional small fraction of
isolated particles as a function of the value of this fraction. As shown
from the presented set of curves, the spectrum of such an ensemble becomes
very sensitive to the presence of even a small amount of free particles. The
addition of 1/50 to 1/10 parts of free particles (in relation to the number
of particles comprising a fractal aggregate) leads to a gradual shift of the
principal spectral maximum towards the long-wavelength range. This fact
underlines the resemblance of these curves to the experimental dependencies.
However, a gradual reduction in the contrast of a secondary maximum is
observed in this case.

Spectral dependencies obtained in this work, along with data on the FPSD
\cite{9}  and the absorption spectra of aggregated silver colloids, present
a unique opportunity to quantitatively compare current experimental data
with the results of calculations performed by the proposed procedure.

\begin{figure*}[!h]
\begin{center}
\includegraphics[width=0.5\textwidth]{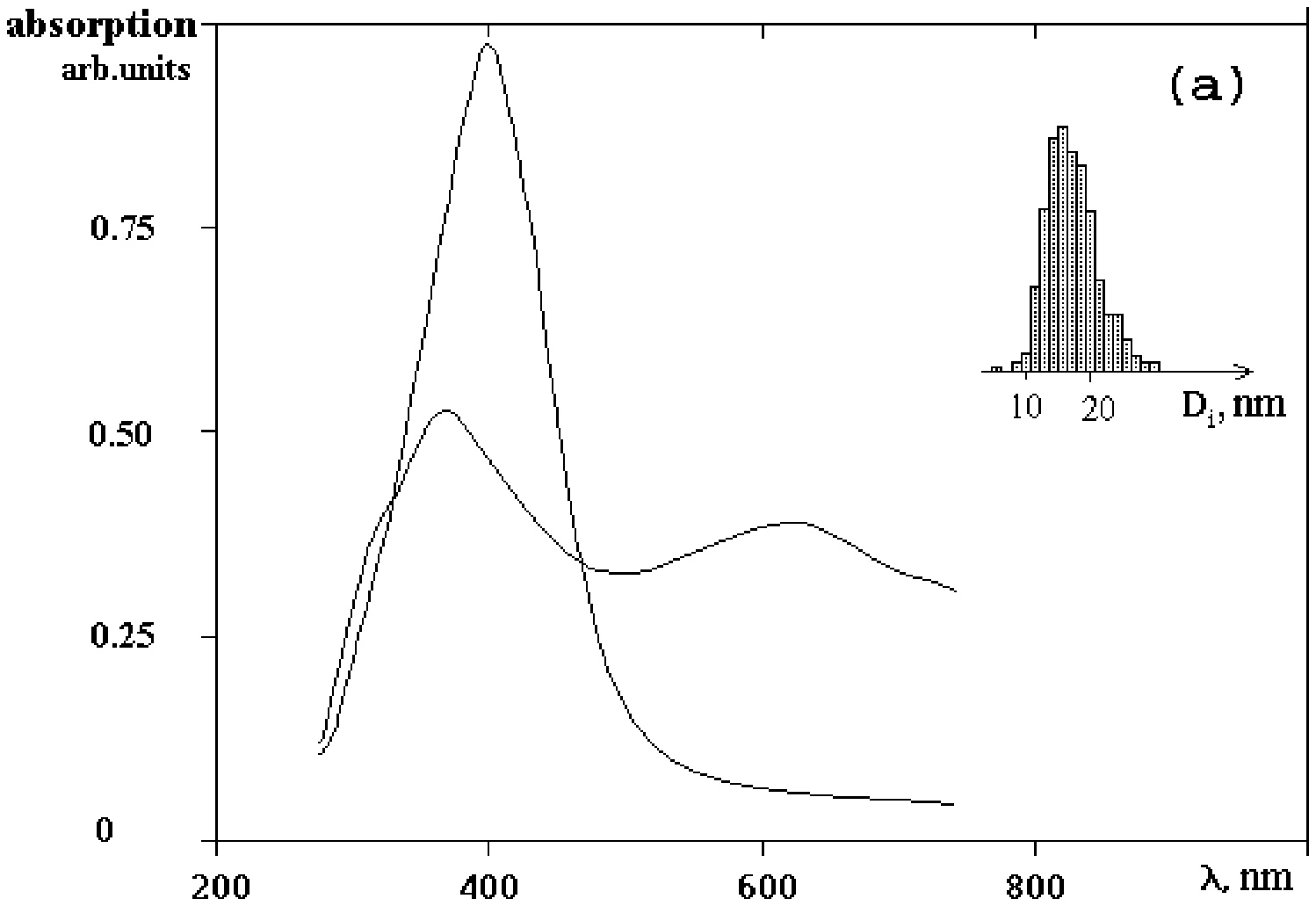}
\vspace{1mm}
\includegraphics[width=0.5\textwidth]{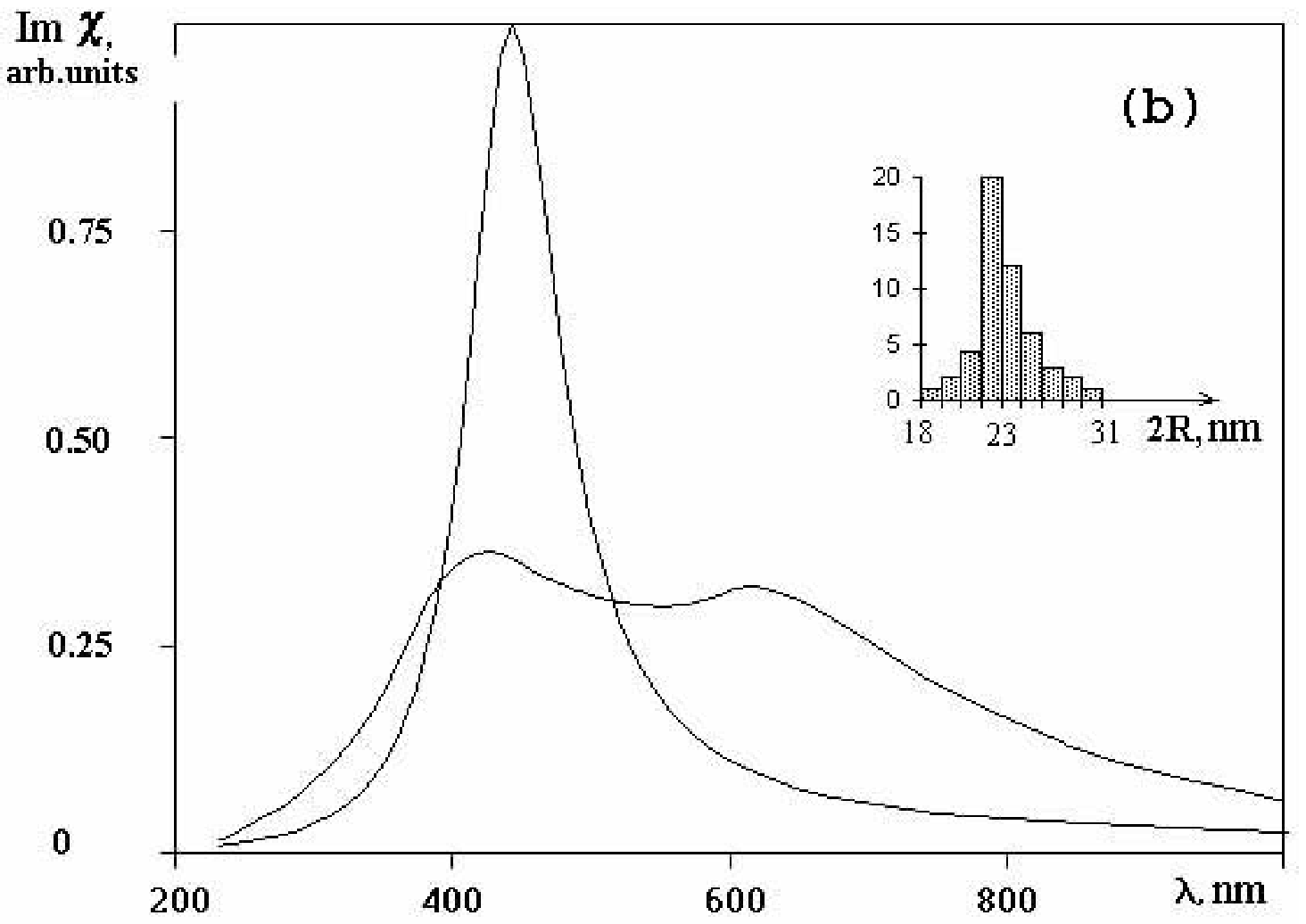}
\caption[recon] {\label{f15} Observed and calculated absorption
spectra of silver hydrosols. (a) Observed absorption spectra at
(1) the initial and (2) developed stages of aggregation \cite{9}
and corresponding size-distribution function; (b)  Based on the
general theory of OPFC, calculated absorption spectra for the
size-distribution function similar to that reported in \cite{9}.
In the case of (a), $D_i$ is the diameter of the metallic core of
the particles.}
\end{center}
\end{figure*}

Figure 15 shows the absorption spectra of real silver colloids with
corresponding FPSD \cite{9}, and the results of calculations of these
spectra  based on the theory of OPFC with the FPSD are closest to those
reported in \cite{9} ($2R_m  = 16 - 17$ nm). As is seen from the comparison
of these data, the most similar pattern of spectra, when the position of a
low-energy maximum coincides with its experimental value, is accomplished
for the FPSD with particle sizes somewhat larger ($2R_m\approx21$ nm) than
those reported in \cite{9}. The revealed difference is explained by the fact
that, during the study of the FPSD, Heard {\it et al.}  \cite{9} only took
into account the size of the metallic core of particles. Meanwhile, the
polar surfactant was used to increase the aggregation stability of colloids
to some extent. We believe that, to explain the discrepancy between these
results, we should take into account the speculations reported in Sec. 4.2
and, in particular, expression (\ref{4.2}). Hence, the discrepancy indicated
can be explained by the fact that the adsorption layer of this colloid
contains a surfactant. This should result in an increase in the effective
size of particles comprising the aggregates, in a slight variation in the
FPSD pattern reported in this work and, hence, in the average distances
between the nearest particles and, consequently, in the spectrum patterns.
Let us take advantage of the data on the thickness $L$ of the adsorption
layers of the particles of aggregated sols of some preparations of colloidal
silver \cite{42}, as well as on the microimages of fractal aggregates
\cite{9,42}, in order to estimate $L$ by the value of the spacings between
the nearest particles in the aggregate. On average, these values are equal
to $2 — 4$ nm (with no correction for the dehydration of surfactant/polymer
molecules). This implies that the particle diameter used in the calculations
should be increased by at least $2 - 4$ nm (without accounting for the
$L(2R_i )$ function).

All the calculated dependencies of the absorption spectra mentioned above
(Figs. 1, 8-15) were obtained within the framework of a two-level model
describing the particle dipole polarizability [see formula  (\ref{3.12})].
Figure 14 shows the set of curves of the absorption spectra of aggregated
colloids, which was obtained using a full-length expression for the dipole
polarizability with allowance for the spectral dependence of the
permittivity of the particles comprising the matter and the permittivity of
the dispersion medium. This dependence was accounted  for within the
framework of the Drude model [see formulae (\ref{3.14}-\ref{3.17})]. This
allows us to estimate the role of this factor in the formation of  the
absorption contour of an aggregating colloid. The values of specific
parameters of these formulas can be taken, for example, from \cite{32,33}:
$\epsilon_0$ =6 and $\hbar\omega_p$ = 9.1 eV ($\lambda_p$ = 136.1 nm).
However, the value of the damping constant ($\Gamma$) used in our
calculations corresponds to the experimental value of the spectral width of
plasmon resonance at the half-height ($\Delta\lambda$ = 100  nm) in
collargol-based non-aggregated Ag hydrosols. Calculations of the spectra
within the framework of a two-level model were performed for the same value
of $\Gamma$. Variations in this parameter lead to results similar to those
represented in Fig. 10.
\begin{figure*}[!h]
\begin{center}
\includegraphics[width=0.5\textwidth]{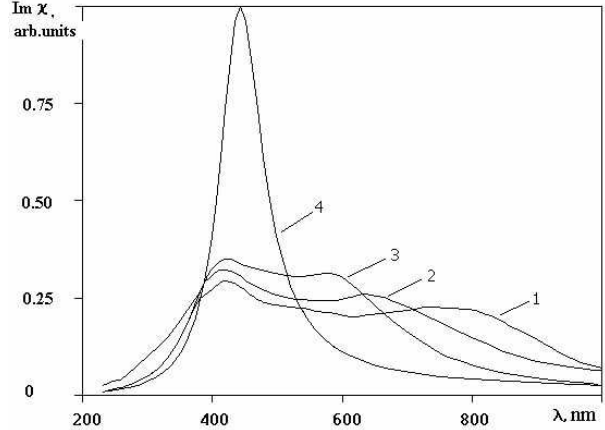}
\caption[recon] {\label{f16} Absorption spectra of fractal
aggregates simulated by using the expression for the particle
dipole polarizability, which accounts for the spectral dependence
of a metal permittivity (the approximation of the Drude model).
The size distribution corresponds to $2R_m$ of (1) 18.2,  (2)
23.1, (3) 26.4 nm (see the similar curves 2, 5, and 7 in Fig. 12),
and (4) spectrum of isolated particles.}
\end{center}
\end{figure*}
While performing the calculations represented in Fig. 16, we used
the same pattern  of FPSD, as for the plots shown in Fig. 12 (with
a two-level model). When comparing the set of curves, we observed
the resemblance and conservation of the main spectral features in
the optical range. The appearance of a secondary maximum in these
spectra also confirms the version of its statistical origin, and
it further explains why this feature cannot account for the
spectral behavior of the optical constants of silver, because only
monotonic variations in these parameters are observed in the
optical range \cite{48}. The results obtained verify the use of
the approximation of a two-level model to describe the particle
dipole polarizability during the calculations of the absorption
spectra of silver colloids. This approximation allows us to
successfully monitor the evolution of sol spectra, with an account
for the contribution of only one selected resonance in the absence
of other close resonances, which is directly applicable to silver
sols in view of their spectral features.

A further increase in the accuracy of the calculations of the optical
spectra of colloids and their fit to experimental results is determined by
the account for such factors as the contribution of the inter-band
transitions to the dipole polarizability of particles, the real spectral
dependence of optical constants for the comprising particles and their size
dependence, the real thickness of the adsorption layers $L(R_i )$, as well
as for scattering effects and for some other factors. It is worth noting
that the possibility of applying the theory of OPFC to interpret absorption
spectra of silver hydrosols is mentioned in  \cite{42}.

Note also that the spectral regularities revealed in this work are virtually
independent of the mechanism of aggregation and the presence of an electric
charge on the particles. The average values of the fractal dimension of
generated aggregates, for which the calculations of optical spectra were
performed, fall within the range $1.65 - 1.78$, and thus only slightly
affect the pattern of the spectral contour.  The spectral regularities are
independent of the number of aggregating particles that was found when
increasing this number by tenfold.

 \subsection{Spectral determination of the aggregation degree}

Here we employ the concept that the structural and optical properties of
fractal aggregates correlate strongly. Hence, we can propose a new method
for determining the aggregation state of the colloids. It can be widely used
as an indirect express-method for monitoring of the aggregation state.

To describe quantitatively the degree of aggregation for silver hydrosol on
the basis of the pattern of its absorption spectrum $\varepsilon(\lambda)$
(Fig. 17), we introduce the following parameter (hereafter called the degree
of aggregation): $A= A_p/A_0$, where  $A_p=\Delta S$/$\varepsilon_0$. Here
$\Delta S$ is difference of the areas below the absorption profiles for
aggregated and non-aggregated sols in that part of the long-wavelength range
where the first exceeds the second, and $\varepsilon_0$ is absorption
coefficient in the range of unshifted plasmon resonance for the aggregated
colloid. The latter parameter is introduced into the formula in order to
account for a possible decrease in the concentration of the dispersed phase
of a colloid due to the partial deposition of the latter onto the vessel
walls. This effect may cause no change in the shape of the spectral curve,
and its magnitude may vary for different samples. The parameter $A_0$ is the
normalizing factor, which is equal to the degree of hydrosol aggregation
immediately before or after precipitation when the phases separated are
subjected to forced stirring. This factor is equal to the maximal possible
value of $A_p$ for the particular type of hydrosol considered (Fig. 17,
curve 3, which corresponds to the maximal degree of aggregation).
Normalization to $A_0$ is required in order to bring the value of $A$ closer
to unity for the media with strong aggregation. In addition, such
normalization makes the mentioned definition of the degree of aggregation,
to a certain degree, universal with respect to different types of hydrosols.

Practically, calculations can be done with the aid of equation,
$A_p=\sum_{i=1}^n{\Delta\varepsilon_i}/{n\varepsilon_0}$, where $n$ is the
number of points selected to break the $\lambda$-axis into equidistant
segments in the region where the absorption of the aggregated medium (curve
2) exceeds the absorption of the medium containing the isolated particles
(curve 1); $\Delta\varepsilon_i$ is the difference between the absorption
coefficients for curves 1 and 2 at the $i$th point ($i$ = 1, 2, ... $n$);
and $\varepsilon_0$ is the absorption coefficient for curve 2 in the region
of the unshifted plasmon resonance.
\begin{figure*}[!h]
\begin{center}
\includegraphics[width=0.5\textwidth]{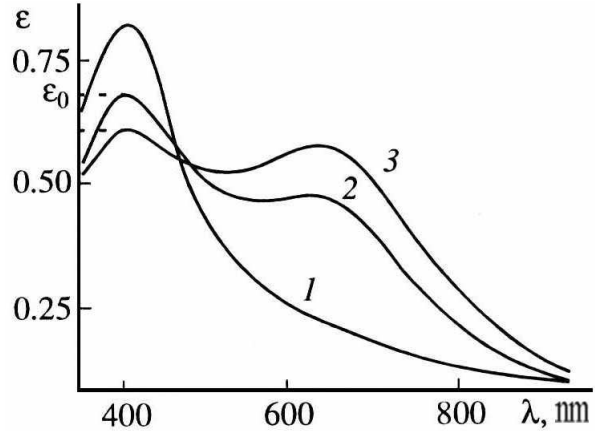}
\caption[recon]{\label{f17} Specific features of the absorption
spectra $\varepsilon (\lambda )$ of silver hydrosols at a specific
degree of aggregation $A$: (1) $A$ = 0  - hydrosol with isolated
particles; (2) $A$ = 0.8 -  intermediate stage of aggregation when
a significant fraction of particles is combined into fractal
aggregates; and (3) $A$ = 1 - all the particles are combined into
fractal aggregates. }
\end{center}
\end{figure*}
Evidently, $A$ is calculated with an accuracy depending on the
selected number of points $n$. In fact, we propose to define $A$
as the difference between the definite integrals taken over the
functions describing curves 2 and 1 at approximately $\lambda  >
440$ nm. In particular, the value of $A$ corresponding to curve 2
(Fig. 17) is equal to 0.8 at $n$ = 8.
\begin{figure*}[!h]
\begin{center}
\includegraphics[width=0.5\textwidth]{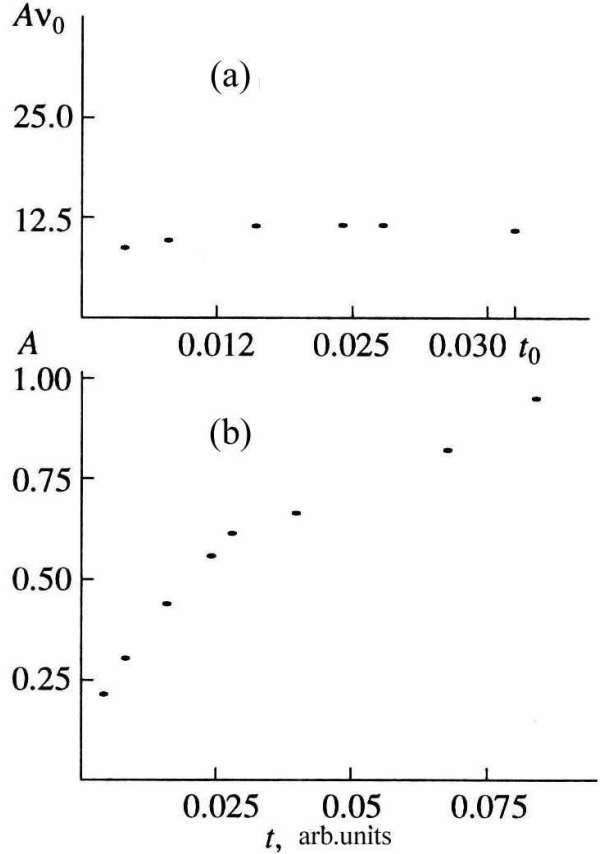}
\caption[recon] {\label{f18} Time dependence of (a) the degree of
aggregation (according to the spectral data) multiplied by the
total number of all particles in the system (including the
composite ones) $A\nu_0$ for the process of spontaneous
aggregation of 50 particles in the time period restricted by the
condition $50> \nu_0>  25$ (numerical simulation) and (b) the
degree of aggregation $A$  at $50 > \nu_0> 1$ (the same model) }
\end{center}
\end{figure*}
The correctness of the method proposed to determine the degree of
aggregation $A$ from the spectral broadening can be verified by the
numerical simulation of particle aggregation under random-walk conditions
while simultaneously  calculating the optical spectra of the forming
aggregates on the basis of the equations from Section 3 and using these
spectra for evaluating $A$. With this, one can perform a comparative
analysis of the time variations of various parameters characterizing the
degree of aggregation in the system of particles. Such parameters can be
introduced in different ways. In particular, the kinetic theory by
Smoluchowski implies that the degree of aggregation is a quantity inversely
proportional to the total number of all the particles that appear in the
coagulated system ($\nu_0^{-1}$ )(see, e.g., \cite{7}).

The parameter $\nu_0$ accounts for both the isolate particles and
associations of two or more particles. It is related to the
initial number of isolated particles $\nu$ by the expression
$\nu_0 = \nu/(1 + t/t_0)$, where $t_0$ is the time required for
$\nu_0$ to decrease to the value $\nu/2$.  The $\nu_0^{-1}\propto
t$ dependence corresponding to this expression holds in the time
interval $0 < t < t_0$ during spontaneous Brownian aggregation of
the particles. This result was repeatedly supported by
experiments. Hence, if our approach to use the spectral data for
determining the degree of aggregation is correct, then both the
parameters $A(t)$ and $\nu_0^{-1}(t)$  should identically describe
the aggregation kinetics, and the product $A\nu_0$ should remain
constant during aggregation in the time interval $0 < t < t_0$ (at
$A \geq 0.1$). Figure 18a represents the time dependence of the
$A\nu_0$ product. This dependence was calculated by the 3-D
simulation of spontaneous Brownian aggregation of fifty particles,
whose optical and geometrical properties corresponded to the
particles of the silver hydrosols examined. We see that, within
the statistical scatter, the value of $A\nu_0$ is constant at $0 <
t < t_0$. This fact supports the correctness of our method of
determining the degree of aggregation and indicates that the
parameter mentioned describes adequately the state of the
dispersed phase of the colloid. According to Fig. 18b, the $A(t)$
dependence (which describes the aggregation kinetics by the
simplest model of Brownian motion) is almost linear in the time
interval $0 < t < t_0$. However, the aggregation rate then
decreases with time due to the formation of large aggregates and
the decrease in their mobility (this effect may disagree with the
experimental results because all physical factors are ignored in
this model except for the kinetic one).

Of course, the proposed definition of the degree of aggregation is somewhat
arbitrary, and one must account for the specific features of different types
of colloids in each particular case. Note that the parameter $A$ reflects
not only the fraction of isolated particles forming the fractal aggregates,
but also the particle packing density in these aggregates, which depends on
the specific features of the structure of the adsorption layer on particles
and is individually determined  for each particular type of sol. We should
also note that the approach proposed to describe the degree of aggregation
is valid only for the disperse systems in which absorption prevails to a
large degree over scattering.

\section{CONCLUSION}

The main results of the work are summarized as follows.

1. It is shown that the effect of the particle size on the optical spectra
of unstable aggregating silver colloids is much more pronounced according
the theory of optical properties of the fractal clusters (OPFC) than in the
Mie theory with an account for the differences in the physical mechanisms of
this effect. The application of the Mie theory leads to qualitative
discrepancies between the calculated dependences and experimental data in
the range of small particle sizes. The main spectral changes that are
observed in typical silver colloids are explained by the theory of OPFC,
even without an account for the effects predicted by the Mie theory. The
effect of dispersion of the particle size within the interval $5 - 30$ nm on
the optical properties of the polydisperse sols, which is the subject of the
Mie theory, is not that considerable. Indeed, the calculated shift of the
resonance wavelength $\Delta\lambda_r (2R_i)$ within the size interval $2R_i
= 5 - 30$ nm makes up only 15 nm, whereas the shift detected in the
aggregated silver hydrosol, with such particle sizes, is larger than 400 nm.

2. An adequate description of the evolution of the optical absorption
spectra of typical aggregating Ag sols, with the aggregating particle sizes
in the range of $5 - 30$ nm, is not possible without account for their
dipole-dipole interaction. This is the main origin of the considerable
spectral broadening, especially for the colloidal structures with the
fractal geometry. Indeed, the distance between geometrical centers of the
nearest particles inside the aggregates is the parameter of the crucial
importance for the OPFC theory.  The main reason for the significant
broadening of the absorption spectra of typical silver colloids (particle
size $5 - 30$ nm) is the assembling of particles of the dispersed phase into
fractal aggregates. There exists a clear and strong correlation between the
degree of particle aggregation and the aggregate structure on the one hand
and the shape of optical spectra on the other hand.

3. The shape of the absorption spectrum of the sole in the developed stage
of aggregation is strongly dependent on the distribution of the
aggregate-comprising particles over their sizes. Indeed, individual features
of absorption spectra of various silver colloids are explained by the
differences in the form of this function, with account for both the sizes of
the metallic core of particles and the thickness of the ionic and polymer
component of their adsorption layers.

4. According to the OPFC theory, the aggregation of Ag particles into
fractal structures gives rise to a giant broadening of the long-wavelength
wing of the absorption spectra which may become commensurable with the
magnitude of the resonant frequency itself. Corresponding shifts of the
resonance are attributed to a relatively narrow range of the distances
between the particles, while minimum possible magnitudes of these distances,
$r_{ij}^{min} = R_i +R_j$, are determined by the sizes of the aggregating
particles, $2R_i$ and $2R_j$. The OPFC theory predicts a strong (power)
dependence of the frequency shift on the inverse particle size, i.e.,
\hbox{$\Delta\omega_r\propto(R_i+R_j)^{-3}$}, whereas the Mie theory
predicts a near-linear dependence on this size for the Ag sols (at least in
the size interval  $2R_i = 20-100$ nm, with no significant deviation from
this dependence in the interval  $5 - 20$ nm). Therefore, the Mie theory
predicts a qualitatively different (inverse) dependence as compared with the
OPFC theory.

5. The appearance of an additional maximum in the long-wavelength wing of
the absorption spectra of some colloids occurring at the stage of developed
aggregation has no relationship to the displayed and exhibition of
collective optical resonance of a specific nature, which is  unrelated to
the excitation of the surface plasmon. The appearance of this maximum is
attributed to the existence of the statistically-dominating values of
geometrical parameters in an ensemble: particle size $2R_m$ and distances
($R_i +R_j$) between contacting neighbor particles in aggregates
corresponding to this size. This results in an increase in the spectral
density of the surface plasmon  resonances within a certain range of
wavelengths. In this case, this spectral feature is exhibited in sols with a
sufficiently narrow ($5 - 25$ nm) range of effective particle sizes.

6. The position and shape of the secondary maximum in the absorption
spectrum are determined by the value of the dominant distances between
neighboring particles in aggregates, which depend on $2R_m$.  The lower the
degree of sol polydispersity, the higher is the contrast of the secondary
spectral maximum. The contrast of this maximum decreases with the increase
of the homogeneous absorption spectral width by the isolated particles.
Furthermore, even a small ($1/50 - 1/10$) fraction of non-aggregated
particles decreases the contrast.

7. The absence of two maxima in the absorption spectra of some silver
colloids can be explained by the presence of an excess number of small
particles in a colloid (with account for the thickness of their adsorption
layer) and its strong polydispersity, thus resulting in a sharp decrease in
the contrast of the secondary spectral maximum.

8. In general, two factors play an important role in the formation of the
long-wavelength absorption spectral wing: fractal geometry of the
aggregates, which determines the monotonic decrease of absorption with
increase of wavelength, and the existence of the dominant particle size in
the polydisperse mixture of the aggregate-comprising particles, which leads
to breaking such monotonic dependence and to the appearance of the
additional spectral maximum.

9. The concepts presented in this work remain valid not only for silver
colloids, but can also be extended to colloids of other metals, including
gold, with an account for spectral dependencies of its optical constants.

\section{ACKNOWLEDGMENTS}
The authors are grateful to V. P. Safonov and V. M. Shalaev for useful
discussions related with this research. AKP and TFG thank the U. S. National
Research Council - National Academy of Sciences for support of this research
in part through the International Collaboration in Basic Science and
Engineering program.


\begin{thebibliography}{100}
\bibitem{1} Karpov S.V., Popov A.K., Slabko V.V., {\it et al.}, Colloid J., 1995, vol.
57, no. 2, p. 199.

\bibitem{2}  Mie G., Ann. Phys., 1908, vol. 25, p. 377.

\bibitem{3}  van de Hulst, H.C., Light Scattering by Small
Particles, New York: Wiley, 1957. Translated under the title Rasseyanie
sveta malymi chastitsami, Moscow: Inostrannaya Literatura, 1961.

\bibitem{4}  Wiegel E., Z.Phys. 1954. vol. 136, p.642.

\bibitem{5}  Skillman D.C. and Berry C.R., J. Opt. Soc. Am. 1973,
vol. 63, no. 6, p. 707.

\bibitem{6}  Skillman D.C. and Berry C.R., J. Chem. Phys., 1968,
vol. 48. no. 7, p. 3297; Kreibig U. and Fragstein C.V., Z. Phys., 1969, vol.
224, no. 4, p. 307.

\bibitem{7}Voyutskii S.S., Textbook of Colloid
Chemistry (Kurs kolloidnoi khimii), Moscow: Khimiya, 1976.

\bibitem{8}  Bohren C.F. and Huffman,
D.R., Absorption and Scattering of Light by Small Particles, New York:
Wiley, 1983. Translated under the title Pogloshchenie i rasseyanie sveta
malymi chastitsami, Moscow: Mir, 1986.

\bibitem{9} Heard S.M., Griezer F., Barrachlough C.G., and Sanders J.V., J. Colloid.
Interface Sci.. 1983, vol. 93, no. 2, p. 545.

\bibitem{10} Petrov Yu.L,  Clusters and Small Particles (Klastery i malye
 chastitsy), Moscow: Nauka, 1986.

\bibitem{11} Xu M. and Digham M.J., J. Chem. Phys. 1992. vol. 96, no. 5. p. 3370.

\bibitem{12} Shalaev V.M. and Stockman M.I., Zh. Eksp. Teor. Fiz. 1987, vol. 94.
p. 107 [Sov. Phys. JETP 1987, vol. 67, p. 287].

\bibitem{13} Markel V.A., Muratov L.S., and Stockman M.I., Zh. Eksp. Teor.
Fiz., 1990, vol. 92, p. 819 [Sov. Phys. JETP 1990, vol. 71, p. 455]; Markel
V.A., Muratov L.S., Stockman M.I., and George T.F., Phys. Rev. B,  1991,
vol.  43, p. 8183.

\bibitem{14} Zsigmondy R., Kolloidchemie: ein Lehrbuch, Leipzig: O. Spamer, 1925. Translated
under the title Kolloidnava khimiya, Kiev: UNIS, 1931.

\bibitem{15} Taleb A., Petit C., and Pileni M.P., J. Phys. Chem. B. 1998. vol. 102. no.
12. p. 2214.

\bibitem{16} Lu A.H., Lu G.H., Kessinsjer A.M., and
Foss C.A., J. Phys. Chem., 1997, vol. 101, no. 45, p. 9139.


\bibitem{17} Bruning J.H. and Lo Y.T., IEEE Trans. Antennas
Propog. 1971, vol. AP-19, p. 378, p.391.

\bibitem{18} Borgese F., Denti P., Toscano G. and Sindoni O.I.,
Appl. Opt., 1979. vol. 18. no. 1, p. 116.

\bibitem{19} Gerardy J.M. and Ausloos M., Phys. Rev. B., 1982. vol. 25, no. 6.
 p. 4204.

\bibitem{20} Mackowski D.W., J. Opt. Soc. Am. A, 1994, vol. 11, no. 11. p. 2851.

\bibitem{21} Fuller K., J. Opt. Soc. Am. A, 1994, vol. 11, no. 12, p. 3251.

\bibitem{22} Xu Y.-L., Appl. Opt., 1995, vol. 34. no. 21, p. 4573.

\bibitem{23} Purcell E.M. and Pennypacker C.R., Astrophys. J., 1973, vol. 186, no. 2, p.795.

 \bibitem{24} Ravey J.-C.,  J. Colloid Interface Sci., 1974,
 vol. 46,  p.139.

 \bibitem{25} Jones A.R., Proc. R. Soc. Lond. A., 1979, vol. 366,
 p. 111.

 \bibitem{26} Chiapetta P.,  J. Phys. A: Math. Gen., 1980, vol. 13,
 p. 2101.

 \bibitem{27} Draine B.T., Astrophys. J., 1988, vol. 333, no. 2,  p.848.

 \bibitem{28} Draine B.T. and Flatau P.J., J. Opt. Soc. Am. A, 1994, vol. 11, no. 4,
 p. 1491.

 \bibitem{29} Lumme K. and Rahola J., Astrophys. J., 1994, vol. 425, p. 653.

 \bibitem{30} De Voe H., J. Chem. Phys., 1964, vol. 41, no. 1, p. 393.

 \bibitem{31} De Voe H., J. Chem. Phys., 1965, vol. 43, no. 9, p. 3199.


 \bibitem{32} Markel V.A., Shalaev V.M., Stechel E.V., {\it et al.}, Phys. Rev.
 B., 1996, vol. 53, no. 5, p. 2425; Shalaev V.M., Poliakov E.Y., and
Markel V.A., Phys. Rev. B, 1996, vol. 53. no. 5, p. 2437.

 \bibitem{33} Shalaev V.M., Phys. Rep., 1996, vol. 272, p. 61.

 \bibitem{34} Karpov S.V., Popov A.K., Rautian S.G.,\,
 {\it et al.}, JETP  Lett., 1988, vol. 47,  p. 243.

 \bibitem{35} Halperin W.P., Rev. Mod. Phys., 1986, vol. 58, no. 3, p. 533.

\bibitem{36} Nagaev E.L., Usp. Fiz. Nauk, 1992, vol. 162, no. 9, p. 49.

\bibitem{37} Karpov S.V., Bas'ko A.L., Koshelev S.V., {\it et al.}, Colloid J.,
1997, vol. 59, no. 6, p. 765.

\bibitem{38} Karpov S.V., Popov A.K., and Slabko V.V., Pis'ma Zh. Eksp. Teor. Fiz., 1997, vol. 66, no. 2, p. 97 [JETP
Lett., 1997, vol. 66, no. 2. p.106].

\bibitem{39} Butenko A.V., Danilova Yu.E., Karpov S.V., {\it et al.}, Z. Phys. D, 1990,
 vol. 17, p.283.

 \bibitem{40} Safonov V.P., Shalaev V.M., Markel V.A., Danilova Yu.E.,
Lepeshkin N.N., Kim W., and Armstrong R.L.,  Phys. Rev. Lett., 1998, vol.
80, no. 5, p. 1102.

 \bibitem{41} Karpov S.V., Popov A.K., and
Slabko V.V., Izv. RAN (physics), 1996,  vol. 60, p. 43.

 \bibitem{42} Danilova Ye.E. and Safonov V.P., Fractal Review in the
Natural and Applied Sciences: Proceedings of the Third IFIP Working
Conference on Fractals in the Natural and Applied Sciences, Marseille, 1995,
Chapman and Hall, London, 1995, p. 102; Safonov V.P. and Danilova Yu.E. in
Spectral Line Shapes: 12th International Conference on Spectral Line Shapes,
Toronto, AIP Conf. Proc., 1994, vol. 8, p. 363.

 \bibitem{43} Creighton J.A., Blatchford C.G., and Albrecht M.G., Trans.
Faraday Soc., 1979, vol. 75, p. 790.

 \bibitem{44} Marton J.P. and Jordan B.D., Phys. Rev. B, 1977, vol.
15, no. 4. p. 1719.

 \bibitem{45} Barykinskii G.M. and Tuzikov F.V.,
Serebro v meditsine, biologii i tekhnike (Silver in Medicine, Biology, and
Technology), Novosibirsk: Inst. Klinicheskoi Immunologii, Sib. Otd., Ross.
Akad. Med. Nauk, 1996, vol. 5, p. 136.

 \bibitem{46} Gorchakov V.I. and Nagaev E.L., Dokl. Akad. Nauk SSSR.,
1986, vol. 286, no. 27, p. 339.

\bibitem{47} Lidorenko N.S., Chizhik S.P., Gladkikh N.T., et al.
Dokl. Akad. Nauk SSSR, 1981, vol. 258, no. 4, p. 858.

\bibitem{48} Johnson P.B. and Christy R.W., Phys.
Rev. B, 1972, vol. 6, no. 12, p. 4370.

\end{thebibliography}
\end{document}